\documentclass[twoside]{article}

\usepackage[accepted]{aistats2025}
\usepackage[round]{natbib}

\usepackage{graphicx}
\usepackage{hyperref}
\usepackage{float}
\usepackage{mathtools}
\usepackage[T1]{fontenc}
\usepackage{lmodern}
\usepackage{amsmath}
\usepackage{amsfonts}
\usepackage{subcaption}
\usepackage[linesnumbered,ruled,vlined]{algorithm2e}
\usepackage{amssymb}

\begin{document}


\runningauthor{Manan Saxena, Tinghua Chen, Justin D. Silverman}

\twocolumn[

\aistatstitle{Scalable Inference for Bayesian Multinomial Logistic-Normal Dynamic Linear Models}

\aistatsauthor{Manan Saxena \And Tinghua Chen \And  Justin D. Silverman*}

\aistatsaddress{Pennsylvania State University \And Pennsylvania State University \And Pennsylvania State University}]

\begin{abstract}
  Many scientific fields collect longitudinal count compositional data. Each observation is a multivariate count vector, where the total counts are arbitrary, and the information lies in the relative frequency of the counts. Multiple authors have proposed Bayesian Multinomial Logistic-Normal Dynamic Linear Models (MLN-DLMs) as a flexible approach to modeling these data. However,  adoption of these methods has been limited by computational challenges. This article develops an efficient and accurate approach to posterior state estimation, called Fenrir. Our approach relies on a novel algorithm for MAP estimation and an accurate approximation to a key posterior marginal of the model. As there are no equivalent methods against which we can compare, we also develop an optimized Stan implementation of MLN-DLMs. Our experiments suggest that Fenrir can be three orders of magnitude more efficient than Stan and can even be incorporated into larger sampling schemes for joint inference of model hyperparameters. Our methods are made available to the community as a user-friendly software library written in C++ with an R interface.
\end{abstract}

% one sentence for AI STATS
% 
% A scalable and accurate method for posterior state estimation in Multinomial Logistic Normal Dynamic Linear Models with applications to microbiome time series analysis. 

% Keywords
%
% + Bayesian Time Series
% + Multivariate Analysis 
% + Microbiome Data

\section{INTRODUCTION}\label{section: intro}

Many scientific fields collect longitudinal multivariate count data where the total number of counts is arbitrary (e.g., multinomial observations). These data are often called count compositional as the information in the data relates to the relative frequencies of the categories~\citep{silverman2018dynamic}. These data occur frequently in molecular biology~\citep{espinoza2020applications}, microbiome studies~\citep{silverman2018dynamic,joseph2020efficient,aijo2018temporal}, natural language processing~\citep{linderman2015dependent}, biomedicine~\citep{fokianos2003categoricaltimeseries}, and social sciences~\citep{cargnoni1997bayesian}. Although the counting process used to collect these data is often modeled as multinomial, other sources of noise in the system being studied often lead to extra-multinomial variation. While some account for this extra-multinomial variability with multinomial-Dirichlet models~\citep{mosimann1962compound}, multinomial logistic-normal models are often superior, as they can account for both positive and negative covariation between multinomial categories~\citep{aitchison1980logistic, cargnoni1997bayesian,joseph2020efficient,silverman2018dynamic}. Moreover, under suitable transformation (i.e., link function), the logistic-normal is multivariate Gaussian. This facilitates development of latent Gaussian models~\citep{aitchison1982statistical}. Multinomial logistic-normal dynamic linear models are particularly appealing for count compositional time series due to their flexibility and expressiveness~\citep{silverman2018dynamic,joseph2020efficient,cargnoni1997bayesian}.

Dynamic Linear Models (DLM) are a flexible approach to Bayesian time series analysis~\citep{west2006bayesian}. These are linear Gaussian state space models that can perform a wide range of time series decomposition tasks, forecasting, and smoothing. The models can express additive combinations of a wide range of auto-regressive, moving average, dynamic regression, seasonal, and polynomial trend processes~\citep[Chapter 4]{prado2021book}. Multinomial Logistic-Normal DLM (MLN-DLMs) have been successfully applied in multiple fields, including the social sciences~\citep{cargnoni1997bayesian} and, more recently, the study of host-associated microbial communities (microbiomes)~\citep{joseph2020efficient,silverman2018dynamic}. However, computational problems have limited wide-spread adoption of MLN-DLMs. These problems are highlighted by microbiome data which are often high dimensional (tens to hundreds of multinomial categories). In this regime, the lack of conjugacy between the multinomial and the logistic-normal makes many MLN-DLM models intractable. \cite{silverman2022bayesian} recently proposed a theory suggesting that a wide range of Bayesian multinomial logistic-normal models, including MLN-DLMs, MLN Linear Models, and MLN Gaussian Process Regression models, can be efficiently and accurately inferred using a novel sampler with marginal Laplace approximation, the Collapse-Uncollapse (CU) sampler. While their method performs well for the latter two classes, we found it fails for MLN-DLMs due to extreme numerical instability when applied to long time series.

% NOTE Manan, in our response we highlighted FOUR (not 2) extensions beyond Silverman et al 2022. Your writing made it seem like there were only two. 
This article develops efficient and accurate posterior inference methods for a broad family of MLN-DLMs. For brevity, we call our approach \textit{Fenrir}. Fenrir is inspired by the Collapse-Uncollapse (CU) sampler but introduces four key innovations that improve on the MLN-DLMs proposed in \citep{silverman2022bayesian}. First,  Fenrir uses a novel gradient filtering algorithm for \textit{Maximum A Posteriori (MAP)} estimation that is both efficient and numerically stable. Second, Fenrir uses a novel \textit{Debiased Multinomial Dirichlet Bootstrap (DMDB)} for posterior uncertainty quantification that is accurate and far more efficient than the marginal Laplace approximation of \citet{silverman2022bayesian}. Third, we show that Fenrir is efficient enough to be embedded within larger samplers, allowing for joint inference with model hyperparameters. Finally, we extend our models to allow for joint inference across multiple time-series and missing observations. As there are no general-purpose implementations of MLN-DLMs against which we can compare, we also develop an optimized Stan~\citep{carpenter2017stan} implementation of our MLN-DLM model class. We provide optimized and user-friendly software implementations of both methods. Through real and simulated data studies, we find that Fenrir provides accurate posterior estimation while typically being three orders of magnitude more efficient than Stan. 

This article is organized as follows: Section 2 provides an overview of related work. Section 3 presents our proposed approach, highlighting our key contributions. Section 4 demonstrates our approach through simulated and real microbiome data applications. Finally, we conclude with a discussion in Section 5.

\section{RELATED WORK}\label{section: lit-review}

While this work focuses on MLN-DLMs; for completeness, we review the broader inference of Bayesian MLN models. We summarise current approaches into four categories: Hamiltonian Monte Carlo (HMC) and Markov Chain Monte Carlo (MCMC) methods, data augmentation techniques, variational inference methods, and collapse-uncollapse methods. 

\paragraph{Hamiltonian Monte Carlo (HMC) and MCMC Methods:} Early studies of MLN-DLMs used Metropolis-within-Gibbs samplers~\citep{cargnoni1997bayesian}. These methods could only scale to a few multinomial categories~\citep{silverman2022bayesian}. More recently, MLN (and MLN-DLMs) have been inferred using HMC~\citep{grantham2020mimix,aijo2018temporal,silverman2018dynamic} with Stan implementations being particularly popular~\citep{aijo2018temporal,silverman2018dynamic}. Still, these methods either required strong low-rank assumptions~\citep{grantham2020mimix}; could only be applied to a subset of observations at a time~\citep{aijo2018temporal}; or could only scale to 10 multinomial categories and could only handle simple random-walk dynamics~\citep{silverman2018dynamic}. Even with these restrictions, many methods still took hours to infer~\citep{silverman2018dynamic}. 

\paragraph{Data Augmentation Techniques:}  P{\'o}lya-Gamma data augmentation, which introduces P{\'o}lya-Gamma random variables within a Gibbs sampling scheme, has become a popular approach to inferring a wide variety of Bayesian logistic models~\citep{polson2013bayesian}. These methods have even been applied in certain multinomial logistic-normal tasks~\citep{glynn2019bayesian}. However, these methods have a critical limitation: In multinomial logistic-normal models, P{\'o}lya-Gamma random variables cannot be block-updated while maintaining the logistic-normal model form~\citep{linderman2015dependent}. As a result, the number of Gibbs steps required to collect a single posterior sample from these models grows linearly with the number of multinomial categories. In practice, this makes the approach infeasible for the tasks considered in this article. 

\paragraph{Variational Inference:} 
More recently, variational inference has been proposed for MLN models~\citep{joseph2020efficient,silverman2022bayesian}. In fact, \cite{joseph2020efficient} developed a scalable variational approximation to MLN state-space models. However, their method lacks the full generality of the MLN-DLMs class we consider here, i.e., their method only models random-walk dynamics within the state space. Additionally, \cite{silverman2022bayesian} studied variational inference for MLN linear and Gaussian process regression models, concluding that variational inference was more computationally expensive and produced less accurate posterior estimates than Collapse-Uncollapse (CU) methods. 

\paragraph{Collapse-Uncollapse Methods:} 
More recently, \cite{silverman2022bayesian} studied a wide class of Bayesian MLN models that are \textit{Marginally Latent Matrix-T Processes (MLTPs)}. MLTPs are defined by a shared marginal form. These include generalized linear and Gaussian process regression models as well as generalized DLMs (e.g., MLN-DLMs). Their approach leveraged the shared marginal form through a novel \textit{Collapse-Uncollapse (CU) sampler} (see Section \ref{subsection: model-inference} for details). While they showed that the CU sampler alone improved inferential efficiency, they also derived a closed-form Laplace approximation to the Collapse step for MLN models, which yielded the most significant gains.
% Software implementations of their method for MLN linear and Gaussian process regression has become popular for the analysis of microbiome data~\citep{bjork2024longitudinal,holmes2020short,roche2023universal,pereira2022multiomics}.
While \cite{silverman2022bayesian} suggest that their approach could be used for MLN-DLMs, we found their proposed approach is numerically unstable in practice because it requires precomputing the parameters of the matrix-T-process (see Supplementary Section~\ref{section: issue-ltp}). Moreover, while their approach is far more efficient than prior alternatives, the marginal Laplace approximation is still computationally intensive. In this article, we address the problem of numerical stability through a novel filtering-based algorithm for computing the density and gradients without having to ever explicitly compute the parameters as discussed in Section~\ref{section: proposed-method} and Supplementary Section~\ref{section: filsmo}. Moreover, we address the computational complexity of the Laplace approximation through our novel \textit{Debiased Multinomial Dirichlet Bootstrap (DMDB)} approximation. 

\section{PROPOSED METHOD}\label{section: proposed-method}
We consider a \(D \times T\) matrix of multivariate count measurements \(Y\) with elements  \(Y_{dt}\) representing the number of observed counts in category \(d\in \{1,\dots, D\}\) at time point \(t \in \{1, \dots, T\}\). While later sections consider studies that contain multiple time series or time series with missing observations, for simplicity, this section assumes only a single time series with equally spaced observations. 

This article considers the same class of MLN-DLMs introduced in \cite{silverman2022bayesian}:
\begin{align*}
    Y_{.t} &\sim \mathrm{Multinomial}(\pi_{.t}) \\
    \pi_{.t} &= \phi^{-1}(\eta_{.t}) \\
    \eta_t^{T} &= F_{t}^{T} \Theta_t + v_t^{T} \quad \quad\quad v_t \sim \mathcal{N}(0, \gamma_t\Sigma) \\
    \Theta_t &= G_t\Theta_{t-1} + \Omega_t, \quad \Omega_t \sim \mathcal{N}(0, W_t ,\Sigma) \\
    \Theta_0 &\sim \mathcal{N} (M_0, C_0, \Sigma) \\
    \Sigma &\sim \emph{IW} (\Xi,\nu)
\end{align*}
where $\phi$ refers to any log-ratio transformation from the \(D\)-simplex to \((D-1)\)-real space and \(\mathcal{N}(M,A,B)\) denotes a matrix normal distribution. 
For computational efficiency we choose \(\phi=ALR_{D}\) where \(ALR_{D}(x)=(\log \frac{x_{1}}{x_{D}}, \dots, \log \frac{x_{D-1}}{x_{D}})\). However, this choice does not limit the generality of our method as posterior samples can be easily transformed into a wide variety of other log-ratio coordinates systems~\citep[Appendix A.3]{pawlowsky2015book}. The latent log-ratio coordinates, \(\eta_{t}\), are modeled with a multivariate DLM~\citep[Chapter 10]{prado2021book}. These are a flexible family of time series models which, through specification of the terms \(F_{t}, G_{t}, W_{t},\gamma_{t}\), can model a wide variety of polynomial trend, seasonal, and dynamic regression processes, as well as additive combinations of such processes. A thorough review of the capabilities of this model class is provided by \cite[Chapters 4 and 10]{prado2021book}. Here, we simply note
that each of the \(D-1\) dimensions of \(\eta_{t}\) is modeled with \(Q\) state space dimensions, such that the state space \(\Theta_{t}\) is a \(Q\times (D-1)\) matrix. This implies that \(F_{t}\) is a \(Q\)-vector representing how the state-space is related to the log-ratios of the latent multinomial probabilities \(\eta_{t}\); \(G_{t}\) is a \(Q \times Q\) matrix describing the deterministic component of the states temporal evolution; \(W_{t}\) is a \(Q \times Q\) covariance matrix representing covariation between the stochastic component of the states temporal evolution; \(\Sigma\) is a \((D-1) \times (D-1)\) covariance matrix representing covariance between each of the \(D-1\) dimensions of the model; and \(\gamma_{t}\) is a positive-valued scalar (typically set to 1) which allows analysts to model interventions~\cite[Chapter 4]{prado2021book}. Hyperparameters \(\Xi, \nu, M_{0}\) and \(C_{0}\) define the model priors. 
% NOTE Manan, I think this addition is too specific. We could make this point but the way you added it is awkward. You just sort of tacked it on at the end of the paragraph. I think the simplest solution is to just not include this. 
% 
% Note that the Inverse Wishart distribution as a prior on \(\Sigma\) is crucial because it's conjugacy with the latent Gaussian model is necessary for scalability. For relaxing this assumption, a hyperprior could be placed on \(\Xi\) and \(\nu\) but that is beyond the scope of the present article. 

\subsection{Overview of Model Inference}\label{subsection: model-inference}
Our goal is to produce samples from the posterior \(p(\Theta, \Sigma, \eta \mid  Y)\) where \(\Theta=(\Theta_{1}, \dots, \Theta_{T})\) and \(\eta=(\eta_{1}, \dots, \eta_{T})\). We are particularly interested in the posterior marginal \(p(\Theta\mid Y)\), as \(\Theta\) is often of greater scientific interest than \(\eta\) or \(\Sigma\). Our approach is inspired by the CU sampler~\citep{silverman2022bayesian}. We first produce samples from the \textit{collapsed model's} posterior \(p(\eta \mid  Y)\). Then, we \textit{uncollapse} each sample by sampling from the posterior conditional \(p(\Theta, \Sigma \mid  \eta)\). In what follows, we focus on the collapse step (sampling from \(p(\eta \mid Y)\)), as an efficient and exact algorithm for the uncollapse step already exists (see the smoothing recursion provided in Supplementary Section~\ref{section: filsmo}). 

For scalability, we follow \cite{silverman2022bayesian} and seek approximate solutions to the collapsed sampling problem. We develop our solution in two steps: First, we develop an efficient approach for MAP estimation within the collapsed model \(p(\eta \mid  Y)\). Then, we use that MAP estimate to produce samples from an approximate posterior \(q(\eta \mid Y) \approx p(\eta \mid  Y)\). 

\subsection{Efficient MAP Estimation of \texorpdfstring{$\eta$}{eta}}\label{subsection: map-estimation}
% TODO Manan: Equations are part of sentences. They need to end with a period. You made this mistake throughout the manuscript. I will try fix the ones I caught but you need to proofread. 
We obtain MAP estimates of \(\eta\) within the collapsed model:
\begin{equation*}
    \hat{\eta} = \underset{\eta \in \mathbb{R}^{(D-1) \times T}}{\mathrm{argmin}} \left[ -\log p(\eta \mid Y) \right]. 
\end{equation*}

By Bayes rule, the collapsed model can be partitioned into two parts: 
\begin{align*}
    -\log p(\eta \mid Y) \propto &-\underbrace{\sum_{t}\log \mathrm{Multinomial}(Y_{\cdot t} \mid \phi^{-1}(\eta_{\cdot t}))}_{\text{I}} \\ 
&- \underbrace{\log p(\eta)}_{\text{II}}.
\end{align*}
By developing efficient algorithms for calculating each of these terms and their gradients, we can obtain the MAP estimate via first-order optimization methods. 

Term I is straight forward to calculate and the gradients were already provided by \cite{silverman2022bayesian}. Those results are reproduced in Supplementary Section~\ref{section: grad-calc}.

To calculate Term II, we use the fact that
\[\log p(\eta)=\sum_{t}\log p(\eta_{t} \mid  H_{t-1}^{T})\]
where \(H_{t-1}=(\eta_{t-1}, \dots, \eta_{1})\). Following results in \cite{silverman2022bayesian}, each \(p(\eta_{t}\mid  H_{t-1}^{T})\) follows a multivariate \(T\) distribution:
\[p(\eta_{t} \mid  H_{t-1}^{T}) \sim t_{\nu_{t-1}}(f_{t}, q_{t}\Xi_{t-1})\]
where \(\nu_{t-1}\), \(\Xi_{t-1}\), \(f_{t}\), and \(q_{t}\) are given by the multivariate DLM filtering recursion provided in Supplementary Section~\ref{section: filsmo}. Gradients of \(\log p(\eta)\) are therefore the sum of the gradients of each term \(p(\eta_{t} \mid  H_{t-1}^{T})\). In Supplementary Section~\ref{section: grad-calc}, we derive computationally efficient representations of those gradients. 

\subsection{The Debiased Multinomial Dirichlet Bootstrap}\label{subsection: mult-dir}

 Before introducing our approach, we provide relevant context. At least within the microbiome field, researchers have reported success by approximating each observation \(Y_{\cdot t}\) with independent Bayesian Multinomial Dirichlet models~\citep{friedman2012sparcc,fernandes2014unifying,nixon2023sri,nixon2024aldex2}:
\begin{align*}
    Y_{\cdot t} &\sim \text{Multinomial}(\pi_{\cdot t}) \\ 
    \pi_{\cdot t} &\sim \text{Dirichlet}(\alpha)
\end{align*}
where \(\alpha\) is a \(D\)-vector with \(\alpha_{d}>0\). The posterior is given by \(\pi_{\cdot t} \sim \text{Dirichlet}(Y_{\cdot t} + \alpha)\). Those authors use each posterior sample \(\pi^{(s)}=(\pi^{(s)}_{\cdot 1}, \dots, \pi^{(s)}_{\cdot T})\) in subsequent calculations, ultimately summarizing those calculations over \(S\) posterior samples. We call this the Multinomial Dirichlet Bootstrap (MDB). 

Inspired by that approach, we approximate \(\prod_{t} q(\eta_{t}) \approx p(\eta\mid Y)\) where each \(q(\eta_{t})\) is defined by 
\begin{align*}
\pi_{\cdot t} &\sim \text{Dirichlet}\left(\phi^{-1}\left(\hat{\eta}_{\cdot t}\right)\cdot \sum_{d}Y_{dt} + \alpha\right) \\ 
\eta_{\cdot t} &= \phi(\pi_{\cdot t}).
\end{align*}
Here, \(\pi_{\cdot t}\) are the latent parameters of the Multinomial at timepoint \(t\) which are related to the latent log-ratio parameters by  \(\eta_{\cdot t}= \phi(\pi_{\cdot t})\) where \(\phi\) is the \(ALR_{D}\) transformation.

We call this as the \textit{Debiased Multinomial Dirichlet Bootstrap (DMDB)}. The intuition behind this approximation is as follows. In the context of our MLN-DLM model, we expect the posterior mean of the MDB would be biased compared to the true posterior due to the assumed independence between each observation \(Y_{\cdot t}\). In contrast, our MAP estimate \(\hat{\eta}\) does not make this independence assumption and we therefore expect it to be less biased than the posterior mean of the MDB. Moreover, arguments in \cite{silverman2022bayesian} suggest the MAP estimate is a good estimate for the posterior mean \(\mathbb{E}[\eta\mid  Y]\) for MLN-DLMs. As a result, we construct our approximating posterior around this MAP estimate. Our procedure is equivalent to the log-ratio transformation of the MDB but using pseudo-observations, \(\tilde{Y}_{\cdot t}=c_{t} \phi^{-1}(\hat{\eta}_{\cdot t})\), to ensure the maximum of the implied Multinomial Likelihood corresponds to our MAP estimate. We choose the proportionality constant \(c_{t}=\sum_{d}Y_{dt}\) to maintain the strength of evidence of the original data. We use DMDB to approximate the collapsed step of the CU sampler. In later sections we show that the DMDB leads to less biased estimates of the posterior mean than the MDB, supporting our use of the term \textit{Debiased}. 
% NOTE Manan: its bad form to refer to supplementary sections out of order in the main text. I solved that by just not adding the reference and saying we will discuss it later. Not great, but its a Jacky fix. Your writing was also grammatically f***ed ("we provided a figure empirically", are you saying we show the figure empirically? What does that mean? Your just jamming words together). 
% 
% Supplementary Section~\ref{section: simulated-data}, we provide a figure empirically illustrating MDB's bias of the posterior mean compared to Fenrir's DMDB and Stan. 

\subsection{Computational Complexity}\label{subsection: complexity}
Our approach also addresses scalability issues found in prior methods. Even without the challenges of numerical instability, the solution proposed by \cite{silverman2022bayesian} has a computational complexity dominated by the Laplace approximation to the collapsed form. That approximation scales as \(\mathcal{O}([T\times (D-1)]^{3})\) due to the need to decompose a Hessian matrix of size \(T(D-1) \times T(D-1)\). Although their method is much more efficient than prior alternatives, it is still computationally expensive. In contrast, our approach is significantly more efficient. For the collapse step, we use the DMDB approximation, which only requires \(S\) samples from \(T\), \(D\)-dimensional Dirichlet random variables leading to a complexity that scales as \(\mathcal{O}(SDT)\). Our approach is so efficient that the collapse step is no longer rate limiting. Instead, the complexity of our method is dominated by the uncollapse step which scales as \(\mathcal{O}(ST[Q^{3}+(D-1)^{3}])\). 

\subsection{Handling Missing Observations}\label{subsection: missing-obs}
Our approach can easily be extended to handle missing-at-random observations through modifications of the filtering recursion given in Supplementary Section~\ref{section: filsmo}. In these recursions, at each time-point \(t\), the posterior for \(\Theta_{t-1}\) is projected forwards in time to serve as the prior for \(\Theta_{t}\); the data is then used to update the prior to a posterior for \(\Theta_{t}\). When a missing observation is encountered, the posterior for \(\Theta_{t}\) is equal to the prior -- no updating is performed. This simple modification is provided in Supplementary Section~\ref{section: missing-obs}. 

\subsection{Multiple Time Series}\label{subsection: mult-ts}
Our approach can be naturally extended to \(K>1\) time series, with each series potentially having different length \(T_{k}\). We allow each time series to have its own state parameters \(\Theta_{t}\), while sharing information about other parameters (e.g., \(\Sigma\)) between the series.  Operationally, we concatenate these time series and treat them as a single time series with \(T=\sum_{k=1}^{K}T_{k}\) time points, then smooth each time series in isolation. The only exception is that our updates for \(\Xi\) and \(\nu\) during filtering and smoothing share information between the series. Full algorithmic details, along with a graphic depiction of our procedure, are provided in Supplementary Section~\ref{section: multiple-timeseries}. 

\subsection{Hyperparameter Inference}\label{subsection: hyperparameter}

The preceding subsections described an efficient approach to state and covariance estimation \(p(\Theta, \Sigma \mid  Y)\) for any MLN-DLM model. MLN-DLM models are specified by the quadruple \({G_{t}, F_{t}, W_{t}, \gamma_{t}}\). The model class is expanded when hyperparameters within this quadruple are learned from the data. While much of this topic is beyond the scope of the present article, we demonstrate a simple, yet effective approach for joint Bayesian inference of the state variance \(W_{t}\). In later sections, we discuss more complicated, yet likely more efficient approaches to hyperparameter inference. As a demonstration, we assume the state variance is time-invariant and diagonal, \(W_{t}=W=\text{diag}(w_{1}, \dots, w_{Q})\), and extend the MLN-DLM model class by including a prior:
\begin{align*}
    w_{q} &\sim \text{InvGamma}(a_{q},b_{q}). 
\end{align*}
We can generate samples from the posterior \(p(\Theta, \Sigma, \eta, w_{1}, \dots, w_{Q} \mid Y)\) using Gibbs updates detailed in Supplementary Section~\ref{section: hyper-infer}.
% TODO Manan: is this really necessary? I am not sure why you are including it and it sounds awkward AND refers to a supplementary section out of order. 
% 
% In Supplementary Section~\ref{section: local-trend}, we show that Fenrir can model even more complex DLM classes like local trend model and be incorporated into the above structure for hyperparameter inference. 

\subsection{Software Implementation}\label{subsection: software}
We provide software implementations of our inference method for MLN-DLMs, called \textit{Fenrir.} This is implemented as an R package, with the majority of computations implemented as a C++ header library using the Eigen and Boost libraries for efficient linear algebra and random number generation. We use an L-BFGS optimizer (provided by the RcppNumerical library;~\cite{rcppnumerical}) for MAP optimization.

Beyond Fenrir, we also developed an optimized Stan~\citep{carpenter2017stan} implementation of our MLN-DLM models so that we had something against which to compare. Like Fenrir, our Stan implementation is optimized using the CU sampler which improves efficiency and stability of the
% NOTE Manan: be careful about \cite vs. \citep vs. \citet, its a frequent source of typos
method~\citep{silverman2022bayesian}.
As a result, we expect our Stan implementation to be far more efficient than any prior implementation of these models (e.g., \cite{cargnoni1997bayesian,silverman2018dynamic}). For example, our Stan implementation scales to 100 multinomial dimensions whereas others (e.g.,\citet{silverman2018dynamic}) only scale to 10. Stan also implements an L-BFGS optimizer which is nearly identical to the RcppNumerical implementation we use in Fenrir. 

The code for the Fenrir package, the optimized Stan implementation, along with all code required to reproduce the results in the following sections, is available as a GitHub repository (\href{https://github.com/manansaxena/fenrir_paper_code}{github.com/manansaxena/fenrir\_paper\_code}). 

% TODO Manan, this is a really weird way to format a link and, when printed, made it look like we forgot to include the link. See if the above looks better. You didn't include the style files, etc..., so I am not able to build the pdf and look myself. 
% 
% \href{github.com/manansaxena/fenrir_paper_code}{(github link)}.

\section{EXPERIMENTATION AND RESULTS}\label{section: exp-res}
\begin{figure*}[t]
    \centering
    \textbf{Number of Multinomial Categories (D) varying}
    \vspace{0.1cm}
    
    \begin{minipage}{0.32\textwidth}
        \centering
        \begin{subfigure}{\textwidth}
            \centering
            \includegraphics[width=\textwidth]{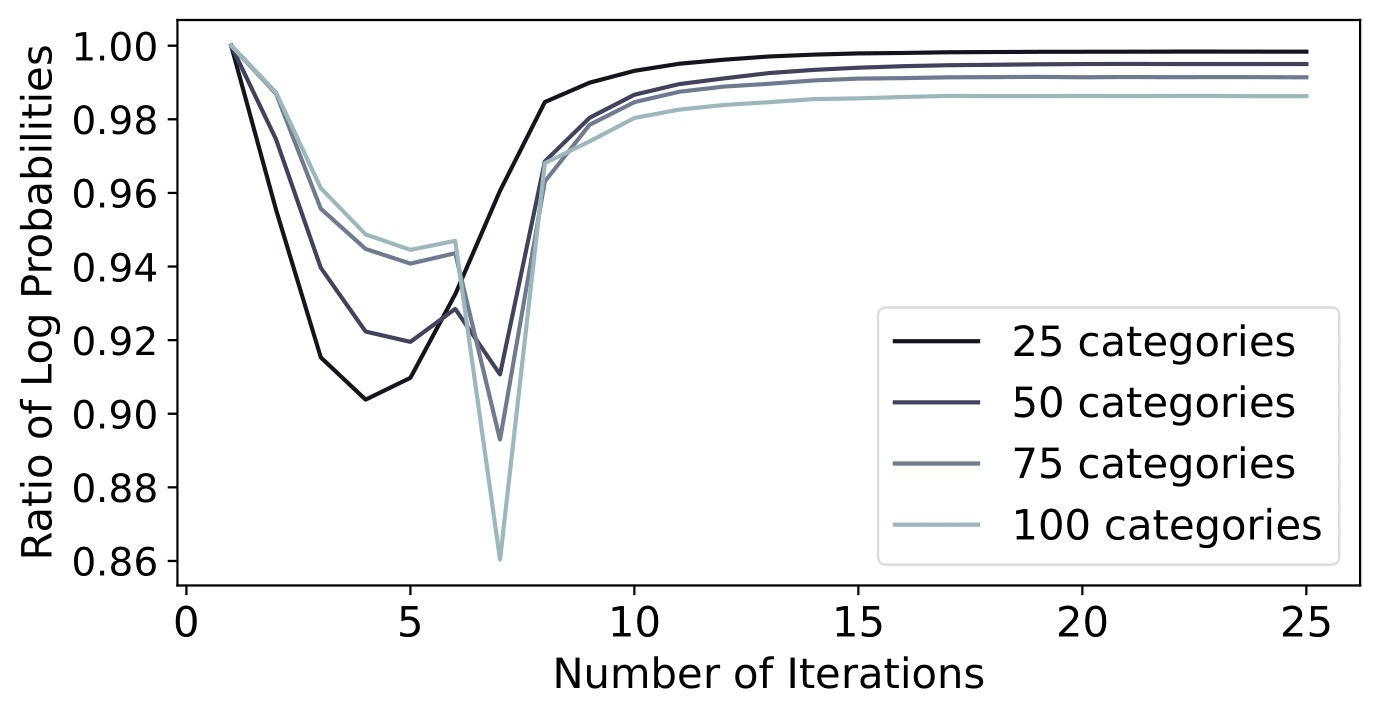}
            \label{fig:D_vary_lp}
        \end{subfigure}
    \end{minipage}
    \hfill
    \begin{minipage}{0.32\textwidth}
        \centering
        \begin{subfigure}{\textwidth}
            \centering
            \includegraphics[width=\textwidth]{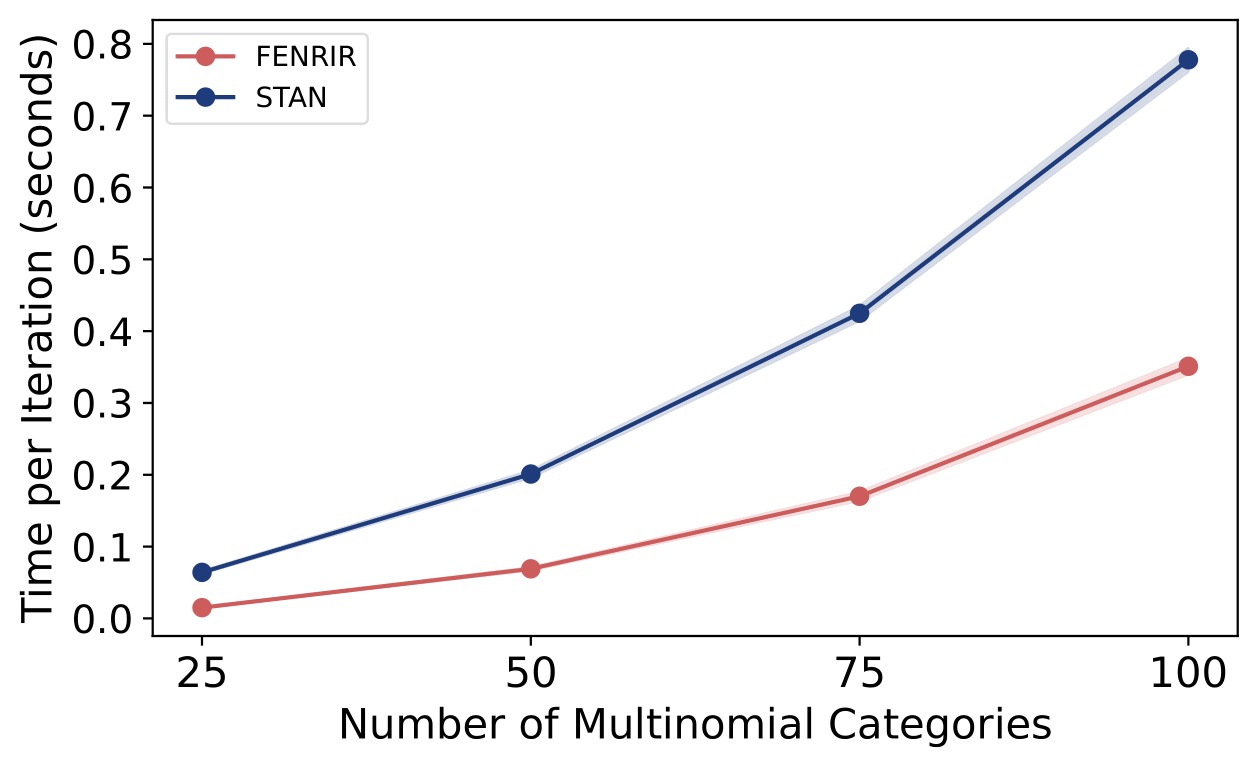}
            \label{fig:D_vary_tpi}
        \end{subfigure}
    \end{minipage}
    \hfill
    \begin{minipage}{0.32\textwidth}
        \centering
        \begin{subfigure}{\textwidth}
            \centering
            \includegraphics[width=\textwidth]{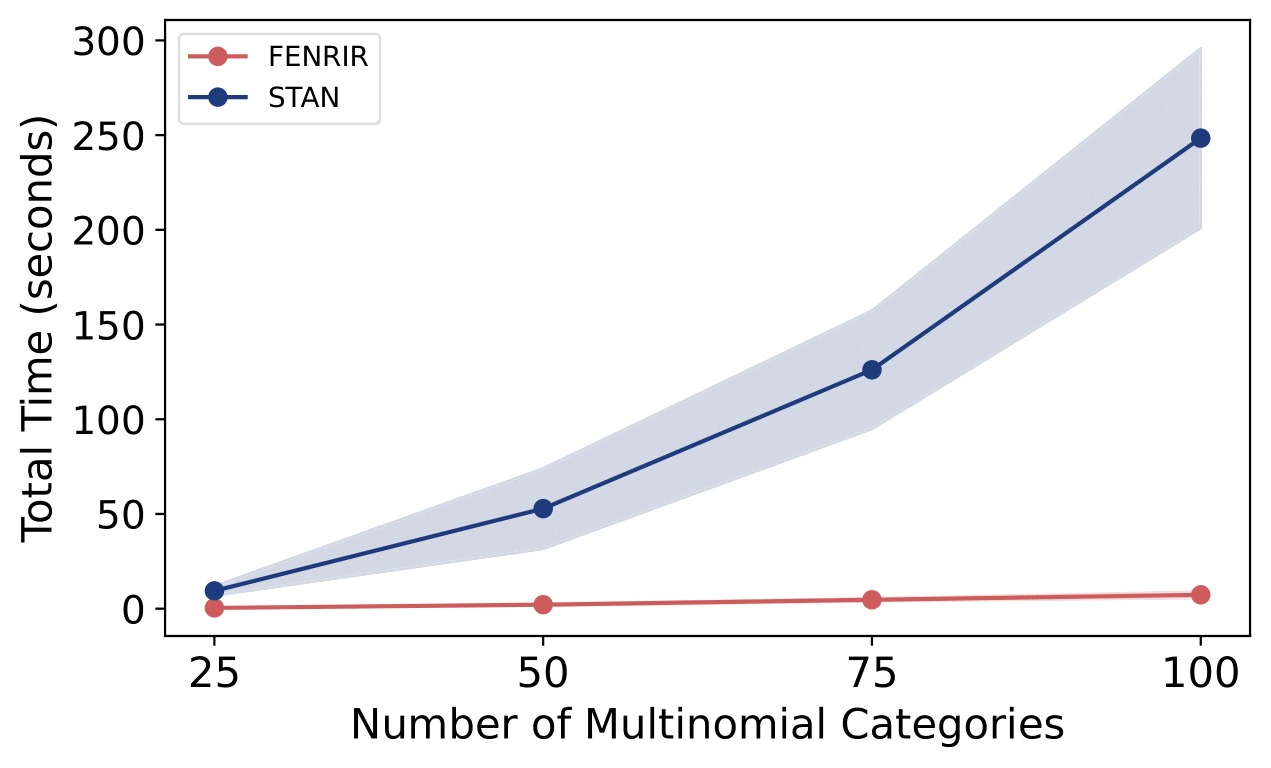}
            \label{fig:D_vary_tt}
        \end{subfigure}
    \end{minipage}

    \vspace{0.1cm}
    \centering
    \textbf{Number of Timepoints (T) varying}
    \vspace{0.1cm}
    
    \begin{minipage}{0.32\textwidth}
        \centering
        \begin{subfigure}{\textwidth}
            \centering
            \includegraphics[width=\textwidth]{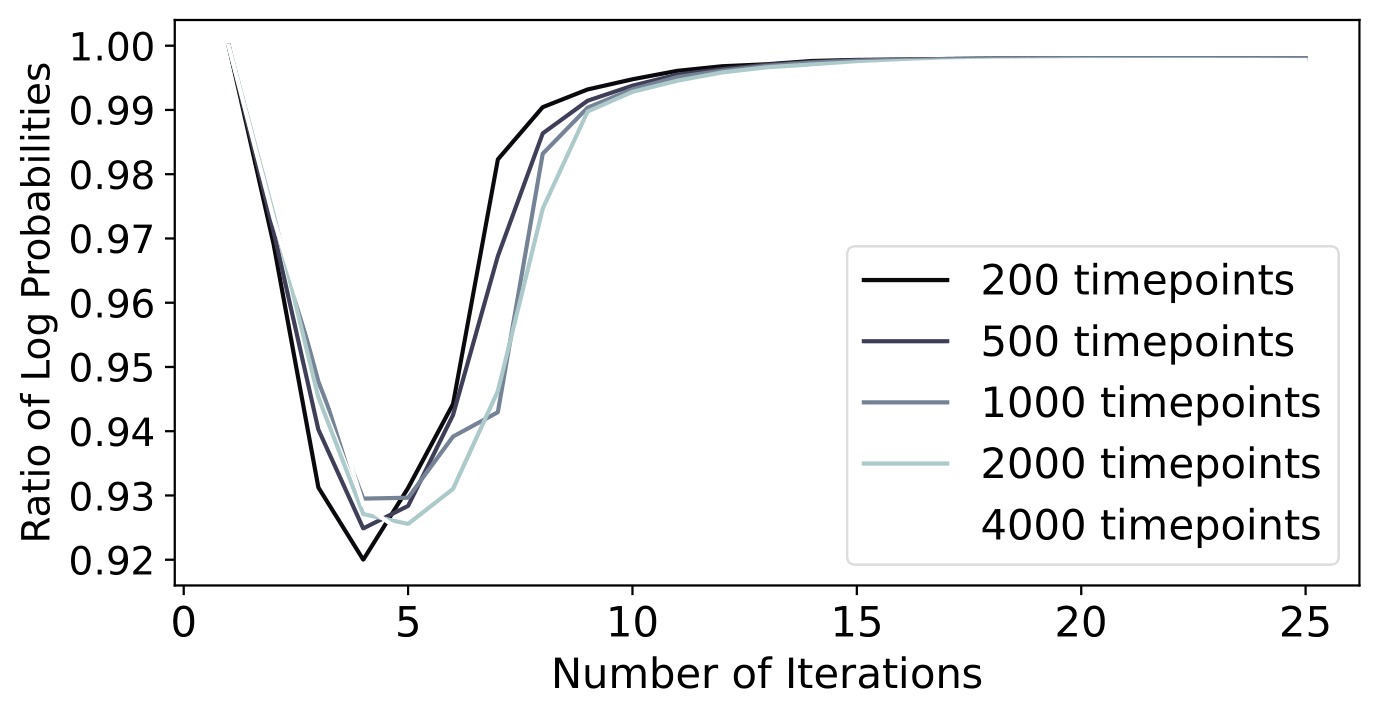}
            \label{fig:N_vary_lp}
        \end{subfigure}
    \end{minipage}
    \hfill
    \begin{minipage}{0.32\textwidth}
        \centering
        \begin{subfigure}{\textwidth}
            \centering
            \includegraphics[width=\textwidth]{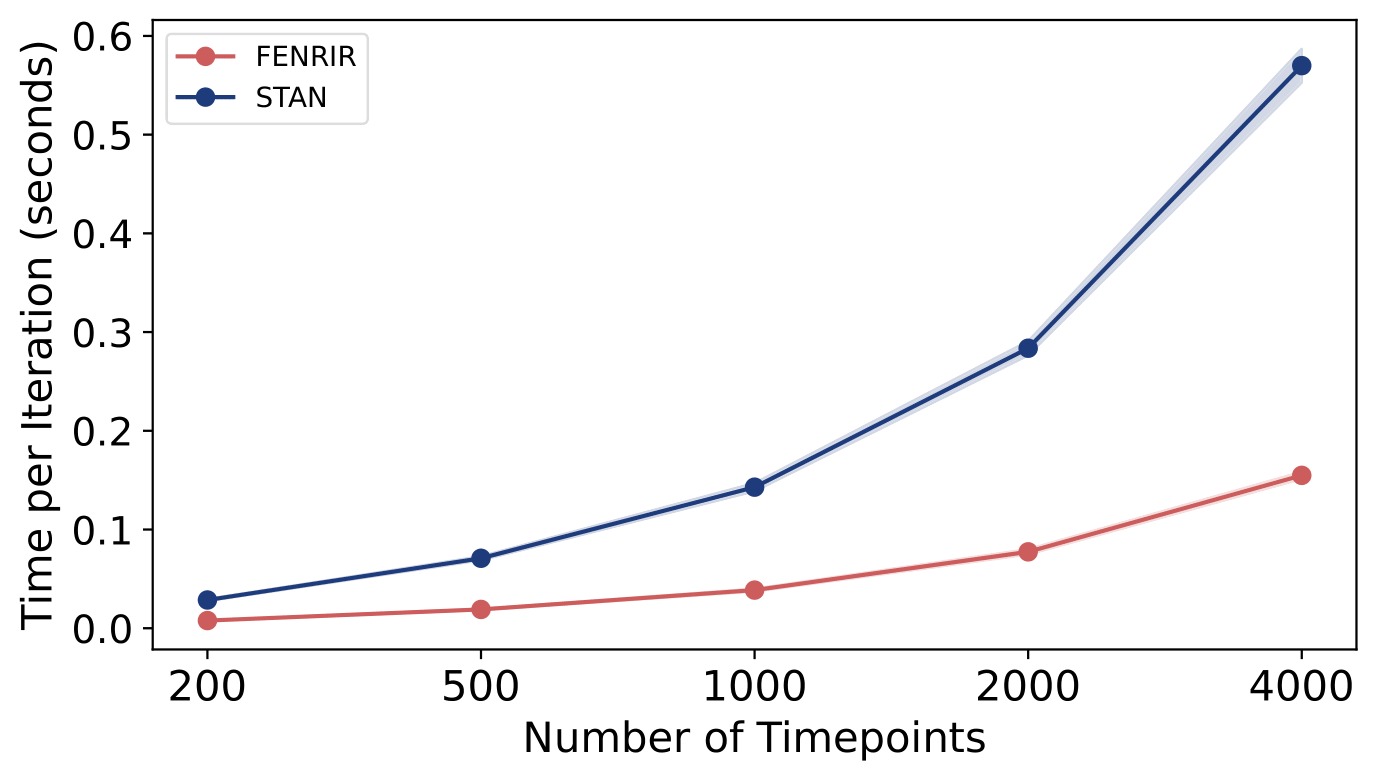}
            \label{fig:N_vary_tpi}
        \end{subfigure}
    \end{minipage}
    \hfill
    \begin{minipage}{0.32\textwidth}
        \centering
        \begin{subfigure}{\textwidth}
            \centering
            \includegraphics[width=\textwidth]{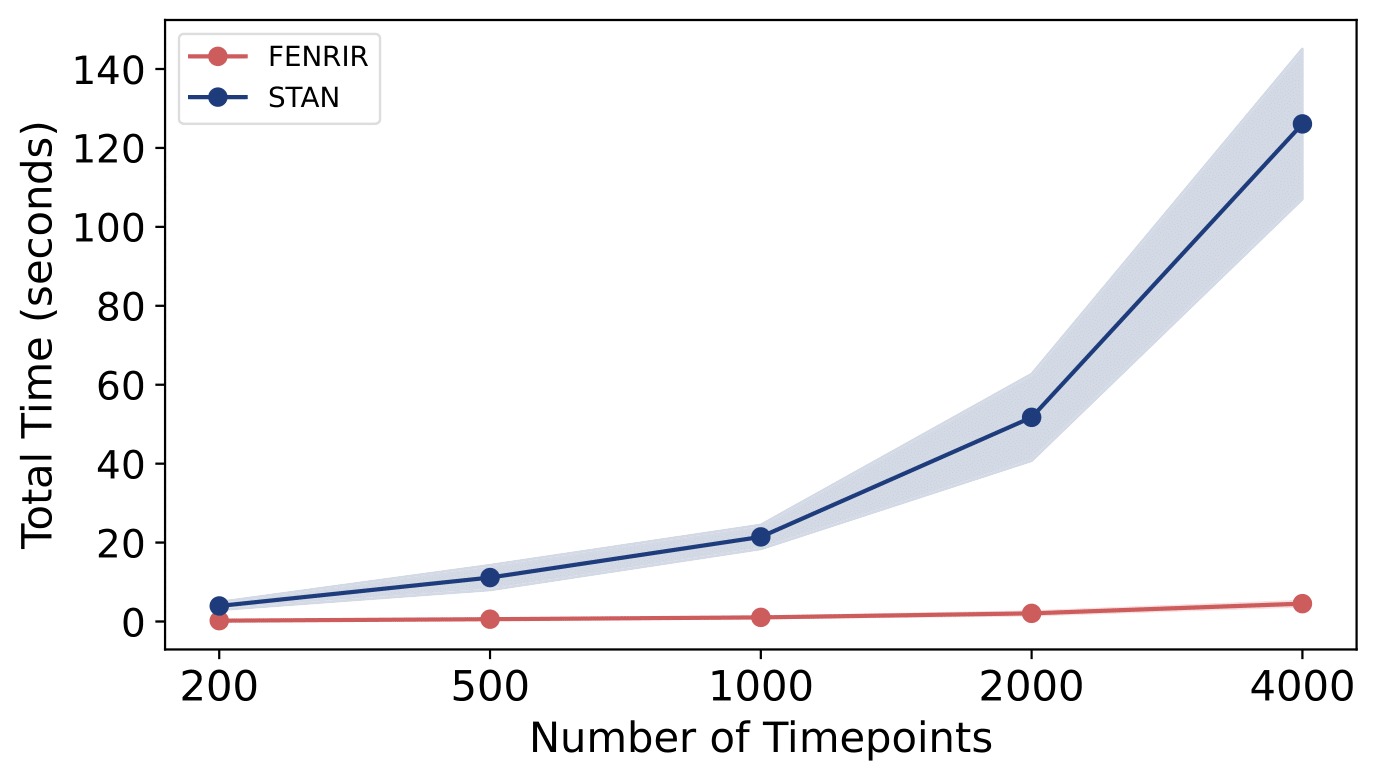}
            \label{fig:N_vary_tt}
        \end{subfigure}
    \end{minipage}
    % \\[2\baselineskip]
    \caption{\textbf{Comparison of optimization results for MAP estimation of \(\eta\) applied to simulated data}. We compared Fenrir to our Stan implementation of MLN-DLMs at various numbers of multinomial categories \(D\) (with \(T\) fixed at 600) and time points \(T\) (with \(D\) fixed at 30). Results represent mean and standard deviations calculated over 10 simulations at each combination of \(D\) and \(T\). Fenrir converges to the same optima as Stan but in fewer iterations (first column), less time per iteration (second column), and substantially lower overall runtime (third column).}
    \label{fig:optimizer-comparison}
\end{figure*}

We compare Fenrir and our Stan implementation (henceforth simply called Stan), in terms of efficiency and accuracy of MAP estimation and uncertainty quantification. In all aspects, we consider Stan the gold-standard in terms of accuracy. All experiments were performed independently on identical hardware, each allocated with 256GB RAM, 16 cores, and restricted to a 48-hour upper limit on wall run-time. Due to Stan's limitations, we restricted our experiments to moderate- and low-dimensional datasets to respect the run-time restriction. 

\subsection{Simulations}\label{subsection: simulation}
To ensure our simulations do not diverge, we simulate a mean reverting random walk:
\begin{align*}
    \Theta_t &= \Theta_{t-1} + \Omega_t, \quad \Omega_t \sim \mathcal{N}(0, 0.45 ,\Sigma) \\ 
    \Sigma &\sim \emph{IW} (I,D+3).
\end{align*}
We simulated multiple time series sharing the parameter \(\Sigma\) between them and introducing randomly missing observations. Full simulation details are provided in Supplementary Section~\ref{section: simulated-data}. Fenrir and Stan were fit to identical data using identical
priors.

\subsubsection{Maximum A Posteriori (MAP) Estimation}\label{subsection: map-sim-res}
We simulated data by varying the number of time points \(T\) and multinomial dimension \(D\). Each simulation was repeated 10 times for each combination of \(T\) and \(D\) to account for random variation in the simulation. The results are summarized in Figure~\ref{fig:optimizer-comparison}. 

Let \(l_{s}(iter)\) and \(l_{f}(iter)\) denote the log-probabilities of the Stan and Fenrir models at a particular iteration, respectively. We compute the ratio \(r=l_{s}(iter)/l_{f}(iter)\) in the first column of Figure~\ref{fig:optimizer-comparison}. As the number of iterations increase, \(r\rightarrow 1\) indicating Fenrir and Stan reach the same optimal. Still, at all iterations, \(r<1\) indicating Fenrir reached the optima in fewer steps. Fenrir also took less time per iteration (second column of Figure~\ref{fig:optimizer-comparison}). Overall this led to substantal decreases in wall run-time (third column of Figure~\ref{fig:optimizer-comparison}). These effects become more dramatic in higher dimensions. For large \(D\) or \(T\), MAP optimization in Fenrir was often 20-30 times faster than Stan while obtaining identical estimates.

We attribute these results to Fenrir's use of closed-form gradients, which likely result in less numerical error compared to Stan's automatic differentiation approach, which involves significantly more computations. Note it is unlikely these effects are due to differences in Fenrir and Stan's L-BFGS implementations as the two are identical save slight differences in stopping criteria. These slight differences could not explain the first and second columns of Figure~\ref{fig:optimizer-comparison}. 

\subsubsection{Uncertainty Quantification}\label{subsesction: uq-sim-res}
For interpretability, we compare the accuracy of uncertainty quantitation in Fenrir and Stan by visualization.  For brevity, we show results for a small simulation \((D=3,T=300)\) in the main text and leave results from a larger \((D=10, T=300)\) simulation to Supplementary Section~\ref{section: simulated-data} as our conclusions were identical in both cases. Additional experimental details and additional simulations are provided in Supplementary Section~\ref{section: simulated-data}.

We compare the computational efficiency of each method using the Number of Effective Samples per Second (NEff/s), a metric recommended by the Stan authors, as it accounts for autocorrelation between samples~\citep{carpenter2017stan}.

\begin{figure}[ht]
    % \vspace{.3in}
    \centering
    \includegraphics[width=\linewidth]{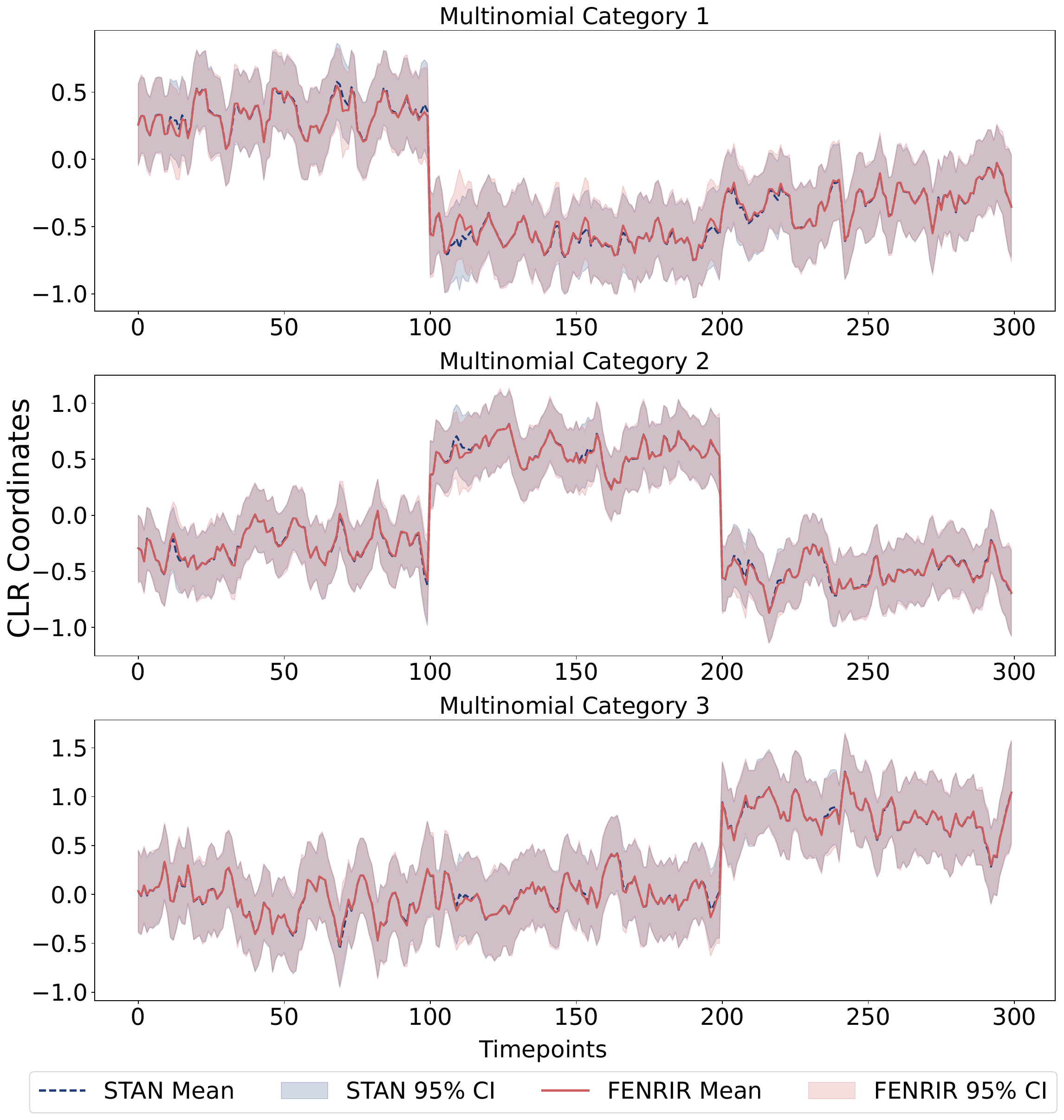}
    % \\[2\baselineskip]
    \caption{\textbf{Posterior mean and credible intervals for \(\Theta\) from Fenrir and our Stan implementation.} The posterior is depicted using centered log-ratio (CLR) coordinates, with each coordinate corresponding to a different multinomial category. \textit{To preempt potential confusion}: both the means and 95\% credible intervals of Stan and Fenrir are plotted but hard to distinguish as there is almost perfect agreement between the estimates.}
    \label{fig:simulated-data-theta}
\end{figure}

Figure~\ref{fig:simulated-data-theta} shows the mean and 95\% credible intervals of the inferred states \(\Theta\) from both Fenrir and Stan. There is almost perfect agreement between the posterior's 95\% intervals and the posterior means estimated by both methods. Yet 
Fenrir produces nearly 800 times more effective samples per second of \(\eta\) than Stan, with Fenrir achieving 12,329 NEff/s compared to Stan's 15.39 NEff/s. To translate this into wall time: in this simulation, Fenrir generates 2000 effective samples (i.e., independent samples) in 0.16 seconds, while Stan requires 129.95 seconds to produce the same number. In short, at least for these simulations, Fenrir has virtually no approximation error while being substantially more efficient.

\subsection{Real Data}\label{subsection: real-data}
We compared Fenrir and Stan on a previously published, finely-sampled, artificial gut microbiome time series. This dataset consists of 4 concurrent time series each consisting of daily measurements over 1 month with high-resolution hourly samples collected during a 5 day period around day 23~\citep{silverman2018dynamic}. This study was previously used to motivate and validate MLN-DLM models~\citep{silverman2018dynamic}. As mentioned in Section~\ref{section: lit-review}, those prior methods took hours to days to obtain accurate posterior estimates and were limited to modeling random walks within the state space for computational tractability. Here, we show that we can reproduce those results in a fraction of the time using Fenrir. Moreover, in Supplementary Section~\ref{section: local-trend} we show that Fenrir can model more complicated state-space dynamics.

We also use this analysis to demonstrate hyperparameter inference. In both Stan and Fenrir, we infer the state variance \(W_{t}=W=\text{diag}(w_{1})\) by adding a \(w_{1} \sim \text{InvGamma}(a_{1},b_{1})\) prior to the MLN-DLM model. In this respect, we expect our simple Gibbs sampler to underperform compared to Stan. Gibbs sampling is often less efficient than methods like HMC or MCMC, where the sampler can move in multiple directions at each iteration (e.g., updating both \(W\) and \(\Theta\)). In contrast, our Gibbs sampler alternates between updates to \(W\) and updates to \(\Theta\).  Still, we find that state estimation using Fenrir is so much more efficient than Stan that it compensates for these limitations, remaining substantially more efficient than our optimized Stan model. In later sections, we discuss how more complicated methods like slice-sampling~\cite{murray2010slicegauss}, could provide further improvements. As in our simulation experiments, both Fenrir and Stan fit identical models: identical data, likelihoods, and priors (see Supplementary Section~\ref{section: microbiome-data} for details).

As in our simulation studies, MAP estimation was substantially more efficient in Fenrir than Stan. At a given value of \(W\), Fenrir produces the same estimate as Stan but in a fraction of the time (see Supplementary Section~\ref{section: microbiome-data}). Overall, Stan took 13.96 seconds for MAP estimation, while Fenrir took only 0.85 seconds. 

\begin{figure}[t]
    \centering
    \includegraphics[width=\linewidth]{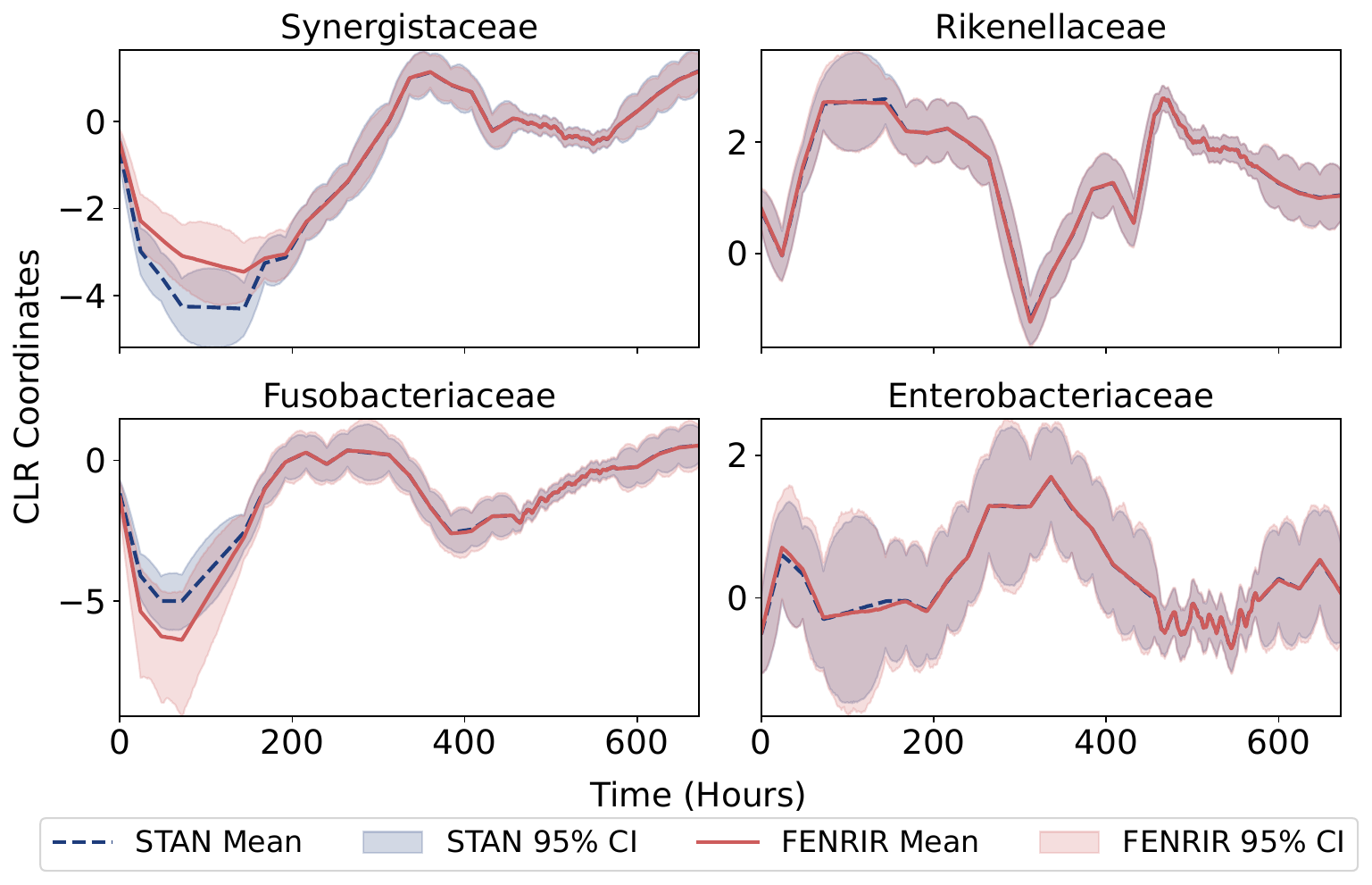}
    % \\[2\baselineskip]
    \caption{\textbf{Posterior mean and credible intervals for state \(\Theta\) of MLN-DLM applied to Artificial Gut Microbiome Data.} For brevity, we show the posterior for \(\Theta\) for four Centered Log-Ratio (CLR) coordinates (four dimensions) in Vessel 2. We highlight the two dimensions with the worst agreement between Fenrir and Stan (Synergistaceae and Fusobacteriaceae) to illustrate worse-case performance.}
    \label{fig:real-data-smoothing-highlights}
\end{figure}

The posterior estimates of \(W\) were nearly identical, with Fenrir estimating a posterior mean and 95\% credible interval of 0.146 (0.132–0.164), and Stan estimating a posterior mean and 95\% credible interval of 0.143 (0.128–0.159). In terms of \(\Theta\), the Fenrir-based Gibbs sampler also produced nearly identical posterior estimates to Stan. For brevity, Figure~\ref{fig:real-data-smoothing-highlights} shows 4 dimensions of the estimated posterior for Vessel~2 (an extended plot is provided in Supplementary Section~\ref{section: microbiome-data}). We purposefully highlight the two dimensions where our Gibbs sampler does the worst (Synergistaceae and Fusobacteriaceae) and two that have received particular attention in previously published analyses (Rikenellaceae and Enterobacteriaceae).

Our Fenrir-based Gibbs sampler was least accurate at the very start of the time series for the taxa Synergistaceae and Fusobacteriaceae (Figure~\ref{fig:real-data-smoothing-highlights}). We attribute this to the abundance of zero counts at the beginning of the time series, when total microbial load in the vessels was low immediately after inoculation~\citep{silverman2018dynamic}. Synergistaceae and Fusobacteriaceae are two of the lowest abundance taxa, which introduces substantial uncertainty regarding their true relative abundance in the presence of so many zero counts. Yet, even in this worst-case scenario, the dynamics inferred by our Fenrir-based Gibbs sampler and Stan are largely in agreement: both methods agree that these two taxa were at extremely low relative abundance, and there is substantial overlap between their posterior 95\% credible intervals. Outside of these 3-4 time points, in these two taxa, there is near-complete agreement between the two methods' posterior means and credible intervals. 

Our Fenrir-based Gibbs sampler produces nearly identical results to Stan for key taxa. As in prior analyses~\citep{silverman2022bayesian,silverman2018dynamic}, the Fenrir-based Gibbs sampler infers a dramatic decrease in the relative abundance of Rikenellaceae following the starvation of Vessels 1 and 2, which occurred between experimental days 11-13. Consistent with prior analyses, we also observe the eventual recovery of the community as Rikenellaceae's relative abundance returns to prior baseline levels. Beyond these large-scale dynamics, our model also captures known fine-scale oscillatory behavior in the Enterobacteriaceae. Similar to previous analyses, we find that this taxon exhibits sub-daily oscillations -- observable when hourly samples were taken during a 5 day period around day 23. 

Overall, the approximate posterior estimated by the Fenrir-based Gibbs sampler was nearly identical to that estimated by Stan. However, 
as expected, posterior samples of \(W\) from the naive Gibbs sampler were more correlated than Stan's (Supplementary Section~\ref{section: microbiome-data}). Despite this, the accelerations provided by Fenrir were enough to overcome this limitation. Overall, the Fenrir-based Gibbs sampler produced effective samples at a rate of approximately 2.5 times Stan (Fenrir: 0.74  NEff/s; Stan 0.31 NEff/s). In the next section, we discuss alternatives to Gibbs sampling that could improve these results.

\section{DISCUSSION}\label{section: discussion}
We have developed efficient and accurate posterior inference for Multinomial Logistic-Normal Dynamic Linear Models (MLN-DLMs). This family of models is flexible and applicable to a wide range of tasks, including forecasting, retrospective inference, and time series decomposition. Beyond microbiome studies, many other fields collect count-compositional time series. These include molecular biology~\citep{espinoza2020applications}, natural language processing~\citep{linderman2015dependent}, biomedicine~\citep{fokianos2003categoricaltimeseries}, and social sciences~\citep{cargnoni1997bayesian}. Our methods may be useful in those fields as well. 

We have developed algorithms for approximate posterior inference. Our experiments show that the approximated posterior can be nearly identical to the true posterior. We suspect the approximation may be good enough to form the basis for an efficient, exact inference algorithm. Methods like importance resampling can refine posterior samples from an approximate model into samples from the exact posterior~\citep[Chapter 6]{prado2021book}. In particular, we suspect sequential particle methods, such as Sequential Importance Resampling or Particle MCMC, will prove particularly useful as they can exploit the structure of time series data. 

This work focuses on state estimation using the Fenrir method described in the main text. While we have demonstrated joint estimation of state parameters and model hyperparameters using a Gibbs sampler, Gibbs schemes are not optimal for this task. In particular, slice sampling can lead to dramatic improvements when inferring covariance hyperparameters in latent Gaussian models~\citep{murray2010slicegauss}. Still, Gibbs schemes are more straightforward and flexible. Remarkably, Fenrir was so efficient as to make even a simple Gibbs sampler almost 2.5 times more efficient than Stan. Overall, Gibbs samplers are likely a good starting point for researchers looking to include Fenrir in more complex models, but ultimately, more specialized methods (e.g., \citep{murray2010slicegauss}) will prove substantially more efficient. 

Our results on real data suggest a limitation of our approach that we suspect may not be addressed effectively by the aforementioned future directions. Based on Figure~\ref{fig:real-data-smoothing-highlights} (and Supplementary Section~\ref{section: simulated-data}), we suspect that the accuracy of our Debiased Multinomial Dirichlet Bootstrap will deteriorate when analyzing time series that are highly sparse (few non-zero counts). This hypothesis is also informed by theoretical results relating to the CU sampler with marginal Laplace approximation~\citep{silverman2022bayesian}, suggesting a similar limitation. While our experimental results suggest our method works well even in the presence of moderate sparsity (e.g., microbiome data), we expect it will perform poorly at levels of sparsity encountered in categorical time series (e.g., natural language processing). As a result, we primarily recommend our method when analyzing multinomial (as opposed to categorical) time series.

\clearpage
\section*{Checklist}

% %%% BEGIN INSTRUCTIONS %%%
% NOTE Justin: The checklist follows the references. For each question, choose your answer from the three possible options: Yes, No, Not Applicable.  You are encouraged to include a justification to your answer, either by referencing the appropriate section of your paper or providing a brief inline description (1-2 sentences). 
% Please do not modify the questions.  Note that the Checklist section does not count towards the page limit. Not including the checklist in the first submission won't result in desk rejection, although in such case we will ask you to upload it during the author response period and include it in camera ready (if accepted).
% %%% END INSTRUCTIONS %%%

\begin{enumerate}
    \item For all models and algorithms presented, check if you include:
    \begin{enumerate}
      \item A clear description of the mathematical setting, assumptions, algorithm, and/or model. [Yes, these details are provided in Section~\ref{section: proposed-method} which includes a full definition of MLN-DLM model and a description of our inference methods. For brevity, that section refers to key results and algorithmic details in the Supplementary Materials.]
      \item An analysis of the properties and complexity (time, space, sample size) of any algorithm. [Yes, it is provided in Section~\ref{subsection: complexity}.]
      \item (Optional) Anonymized source code, with specification of all dependencies, including external libraries. [Yes, Fenrir source code, source code for our Stan implementation, and all code necessary to reproduce the results of this article are provided as a link to github repository.]
    \end{enumerate}

    \item For any theoretical claim, check if you include:
    \begin{enumerate}
      \item Statements of the full set of assumptions of all theoretical results. [Yes]
      \item Complete proofs of all theoretical results. [Yes, key derivations are provided in Supplementary Materials.]
      \item Clear explanations of any assumptions. [Yes]     
    \end{enumerate}

    \item For all figures and tables that present empirical results, check if you include:
    \begin{enumerate}
      \item The code, data, and instructions needed to reproduce the main experimental results (either in the supplementary material or as a URL). [Yes, all code, real microbiome data and a readme file with instructions would be provided as a link to github repository.]
      \item All the training details (e.g., data splits, hyperparameters, how they were chosen). [Yes]
      \item A clear definition of the specific measure or statistics and error bars (e.g., with respect to the random seed after running experiments multiple times). [Yes]
      \item A description of the computing infrastructure used. (e.g., type of GPUs, internal cluster, or cloud provider). [Yes. As described in Section~\ref{section: exp-res}. All experiments were performed independently on identical hardware, each
      allocated with 256GB RAM, 16 cores, and restricted to a 48-hour upper limit on wall run-time.]
    \end{enumerate}
   
    \item If you are using existing assets (e.g., code, data, models) or curating/releasing new assets, check if you include:
    \begin{enumerate}
      \item Citations of the creator If your work uses existing assets. [Yes, We have cited the creators of Stan which is used for creating an optimized implementation and comapres with Fenrir. Furthermore, we have cited the authors of RcppNumerical library used for running LBFGS optimzier in Fenrir. We have also cited the paper where artificial gut data(microbiome data) used for real data analysis was released in Section~\ref{subsection: real-data}.]
      \item The license information of the assets, if applicable. [Yes. Our package Fenrir and paper code follow the GPL-3 license.]
      \item New assets either in the supplementary material or as a URL, if applicable. [Yes, The code of our package Fenrir, our Stan implementation and the code to reproduce results in the paper are shared as a link to github repository.]
      \item Information about consent from data providers/curators. [Not Applicable]
      \item Discussion of sensible content if applicable, e.g., personally identifiable information or offensive content. [Not Applicable]
    \end{enumerate}
   
    \item If you used crowdsourcing or conducted research with human subjects, check if you include:
    \begin{enumerate}
      \item The full text of instructions given to participants and screenshots. [Not Applicable]
      \item Descriptions of potential participant risks, with links to Institutional Review Board (IRB) approvals if applicable. [Not Applicable]
      \item The estimated hourly wage paid to participants and the total amount spent on participant compensation. [Not Applicable]
    \end{enumerate}
\end{enumerate}

\appendix


% Supplementary material: To improve readability, you must use a single-column format for the supplementary material.

\onecolumn
\aistatstitle{Supplementary Material}

\section{ISSUES WITH MARGINALLY LTP FORM FOR MLN-DLM}
\label{section: issue-ltp}

\cite{silverman2022bayesian} propose marginalizing MLN-DLM's to their Latent Matrix T-process form. In brief, they use a recursive filter to marginalize over the state space \(\Theta\) and covariance \(\Sigma\). The problem is that researchers often specify MLN-DLMs with non-stationary state-space models (e.g., random walks or other polynomial trends). We find this leads to numerical instability. While we do not attempt to review their entire proposed approach, we note that it requires calculation of the following matrix \(A\) with elements: 

\begin{equation}
    \begin{aligned}
        A_{t,t-k} &=
            \begin{cases} 
                \gamma_t + F_t^{T} \left[ W_t + \sum_{l=t}^2 G_{t:l} W_{l-1} G^{T}_{l:t} + G_{t:1} C_0 G_{1:t}^{T} \right] F_t & \text{if } k = 0 \\
                F_t^{T} \left[G_{t:t-k+1}\;W_{t-k} + \sum_{l=t-k}^2 G_{t:l}\; W_{l-1}\; G_{l:l-k}^{T} + G_{t:1}\; C_0\; G_{1:t-k}^{T}\right] F_{t-k} & \text{if } k > 0
            \end{cases}
    \end{aligned}
\end{equation}

where \(G_{t:l}\) denotes short hand notation for the product \(G_t \cdots G_l\).

Consider a simple univariate random walk model where \(G=1\). Even in this simple case, for long time series, the term \(\sum_{l=t}^2 G_{t:l} W_{l-1} G^{T}_{l:t}\) explodes as there is linear accumulation of the state variances \(W_{l-1}\). The problem gets worse with more complicated models (e.g., first or second order polynomial models) where the summation leads to quadratic or cubic accumulation. In practice, numerical errors start to dominate the computation and the resultant matrix \(A\). 

The key problem of the \cite{silverman2022bayesian} method is the need to pre-compute the prior over the entire time series -- which requires the matrix \(A\). Instead, our approach avoids this issue. Like the Kalman filter, we only compute the prior one-step ahead of each observation. This stabilizes both our calculation of prior densities (e.g., \(p(\eta)\)) and gradients of those densities. 

\newpage
\section{FILTERING AND SMOOTHING EQUATIONS}\label{section: filsmo}

The following equations define a recursive filter and smoother for the multivariate DLM described in Section~\ref{section: proposed-method} of main text. These equations are reproduced from \cite{silverman2022bayesian} who reproduced them from \cite{prado2021book}. 

\underline{Filtering recursion}:

(1) Posterior at \( t -1 \):
\begin{align*}
    p(\Sigma| H_{t-1}^{T}) &\sim IW(\Xi_{t-1}, \nu_{t-1})\\
    p(\Theta_{t-1} | \Sigma, H_{t-1}^T) &\sim \mathcal{N}(M_{t-1}, C_{t-1}, \Sigma)
\end{align*}

(2) Prior at \( t \):
\begin{align*}
    A_t &= G_tM_{t-1} \\
    R_t &= G_tC_{t-1}G_t^T + W_t \\
    p(\Sigma| H_{t-1}^{T}) &\sim IW(\Xi_{t-1}, \nu_{t-1})\\
    p(\Theta_{t} | \Sigma, H_{t-1}^T) &\sim \mathcal{N}(A_{t}, R_{t}, \Sigma)
\end{align*}

(3) One-step ahead forecast at \( t \):
\begin{align*}
    f_{t}^{T} &= F_{t}^{T}A_{t}\\
    q_{t} &= \gamma_{t} + F_{t}^{T}R_{t}F_{t}\\
    p(\Sigma| H_{t-1}^{T}) &\sim IW(\Xi_{t-1}, \nu_{t-1})\\
    p(\eta_{t}|\Sigma,H_{t-1}^{T}) &\sim \mathcal{N}(f_{t},q_{t}\Sigma)
\end{align*}

(4) Posterior at \( t \):
\begin{align*}
    e_t^{T} &= \eta_t^{T} - f_t^{T} \\
    S_t &= \frac{R_tF_t}{q_t} \\
    M_t &= A_t + S_t e_t^{T} \\
    C_t &= R_t - q_t S_t S_t^{T} \\
    \nu_t &= \nu_{t-1} + 1 \\
    \Xi_t &= \Xi_{t-1} + \frac{e_t e_t^{T}}{q_t}\\
    p(\Sigma | H_{t-1}^{T}) &\sim IW({\Xi}_t, \nu_t) \\
    p(\Theta_{t} | \Sigma, H_{t}^{T}) &\sim \mathcal{N}(M_t, C_t, \Sigma)
\end{align*}

\underline{Smoothing Recursion}:
\begin{enumerate}
    \item Sample $\Sigma \sim IW(\Xi_T, \nu_T)$ and then $\Theta_T \sim \mathcal{N}(M_t, C_t, \Sigma)$.
    \item For each time $t$ from $T - 1$ to $0$, sample $p(\Theta_t | \Theta_{t+1}, H_T^{T})$ from $\mathcal{N}(M_t^*, C_t^*, \Sigma)$ where
    \begin{align*}
        Z_t &= C_t G_{t+1}^{T} R_{t+1}^{-1} \\
        M_t^* &= M_t + Z_t(\Theta_{t+1} - A_{t+1}) \\
        C_t^* &= C_t - Z_t R_{t+1} Z_t^{T}
    \end{align*}
\end{enumerate}

where \(H_{t-1}=(\eta_{t-1}, \dots, \eta_{1})\).

\section{CALCULATION OF GRADIENTS FOR MAP ESTIMATION OF \texorpdfstring{$\eta$}{eta}}\label{section: grad-calc}

As mentioned in the main text Section~\ref{subsection: map-estimation}, we obtain MAP estimates of \(\eta\) within the collapsed model:
\begin{equation*}
    \hat{\eta} = \underset{\eta \in \mathbb{R}^{(D-1) \times T}}{\mathrm{argmin}} \left[ -\log p(\eta \mid Y) \right].
\end{equation*}

By Bayes rule, the collapsed model can be partitioned into two parts: 
\begin{align}\label{eq:2}
    -\log p(\eta \mid Y) \propto &-\underbrace{\sum_{t}\log \mathrm{Multinomial}(Y_{\cdot t} \mid \phi^{-1}(\eta_{\cdot t}))}_{\text{I}} \\ 
&- \underbrace{\log p(\eta)}_{\text{II}}\nonumber.
\end{align}

The gradients for term I (denoted as \(g\) below) were already provided by \cite{silverman2022bayesian} and are given as follows:
\begin{align*}
        g &= \sum_{j=1}^{T} \left( \sum_{i=1}^{D-1} \eta_{ij} Y_{ij} - n_j \log \left( 1 + \sum_{i=1}^{D-1} e^{\eta_{ij}} \right) \right) \nonumber\\
        O &= \exp \eta \nonumber\\
        m &= \mathbf{1}_N + O^T \mathbf{1}_{D-1} \nonumber\\
        \rho &= \text{vec}(O) \oslash \text{vec}\left(\mathbf{1}_{D-1} m^T\right) \\
        n &= \mathbf{1}_D^T Y \nonumber\\
        g &= \text{vec}(\eta)^T \text{vec}(Y_{/D.}) - n \odot \log(m) \nonumber\\
        \frac{dg}{d\text{vec}(\eta)} &= \left( \text{vec}(Y_{/D.}) - \text{vec}(\mathbf{1}_{D-1} n) \odot \rho \right)^T\nonumber
\end{align*}
where  \(\exp X\) and \(\log X\) refers to the element-wise exponentiation and logarithm of a matrix \(X\), \(\odot\) and \(\oslash\) refer to element-wise product and division respectively, and \(Y_{/D.}\) represents the first \(D-1\) rows of the matrix \(Y\).

Term II in equation \ref{eq:2}, as described in the main text's Section~\ref{subsection: map-estimation} follows a multivariate t-distribution with log probability:
\begin{align*} 
    \mathrm{log}\; p(\eta_{t}|H_{t-1}^{T}) \propto \frac{-(\nu_{t-1}+p)}{2}\; \mathrm{log} \left (1+\frac{1}{\nu_{t-1}}(\eta_{t}-f_{t})^{T}(q_{t}\Xi_{t-1})^{-1}(\eta_{t}-f_{t})\right ).
\end{align*}

For getting the gradient, we would like to calculate \(\frac{d\;\mathrm{log} |L|}{d\eta_{t}}\), where \(L = 1+\frac{1}{\nu_{t-1}}(\eta_{t}-f_{t})^{T}(q_{t}\Xi_{t-1})^{-1}(\eta_{t}-f_{t})\):

\begin{align*}
\frac{d\;\mathrm{log} |L|}{d\eta_{t}} = \frac{1}{L}\; \frac{dL}{d\eta_{t}}
\end{align*}

\begin{align*}
    dL &= d\left(\frac{1}{\nu_{t-1}} \left [(\eta_{t} - f_{t})^{T} \; X^{-1} \; (\eta_{t} - f_{t}) \right ]\right) \\
    &= \frac{1}{\nu_{t-1}} \left[ d\eta_{t}^{T}(X^{-1}\eta_{t}-X^{-1} f_{t}) + (\eta_{t}^{T}X^{-1}-f_{t}^{T}X^{-1} )d\eta_{t}
    \right]\\
    &= \frac{1}{\nu_{t-1}} \left[ (\eta_{t}^{T}X^{-1}-f_{t}^{T}X^{-1} )d\eta_{t} + (\eta_{t}^{T}X^{-1}-f_{t}^{T}X^{-1} )d\eta_{t}
    \right]\\
    \frac{dL}{d\eta_{t}} &= \frac{2}{\nu_{t-1}}(\eta_{t}^{T}-f_{t}^{T})X^{-1}
\end{align*}
where $X = (q_{t}\Xi_{t-1})$

\section{MISSING OBSERVATIONS}\label{section: missing-obs}
Missing observations are easy to handle as modifications of the filtering recursion in Section \ref{section: filsmo}. If the \(t\)-th time-point is missing, we simply skip the posterior update and let the prior equal the posterior:
\begin{align*} 
    M_t &= A_t \\ 
    C_t &= R_t \\ 
    \nu_t &= \nu_{t-1} \\ 
    \Xi_t &= \Xi_{t-1} \\ 
    p(\Sigma \mid H_{t-1}^{T}) &\sim \text{IW}(\Xi_t, \nu_t) \\ 
    p(\Theta_t \mid \Sigma, H_t^{T}) &\sim \mathcal{N}(M_t, C_t, \Sigma). 
\end{align*}

\section{MULTIPLE TIME SERIES}\label{section: multiple-timeseries}
The following figure is a graphic depiction of how we handle multiple time-series. In this figure superscripts denote the index of one of \(K\) time-series. Only the filtration over \(\Xi\) and \(\nu\) occur as if all \(K\) time-series were a single long series. For all other parameters filtering and smoothing occur as if the time-series were independent. For example, the state filtration resets to the prior (\(M_0, C_{0}\)) when the filter encounters the start of a new series. Similarly smoothing occurs as if the series were independent. Note that our procedure is invariant to the ordering of the \(K\) series.

\begin{figure}[!h]
    \centering
    \includegraphics[width=0.6\linewidth]{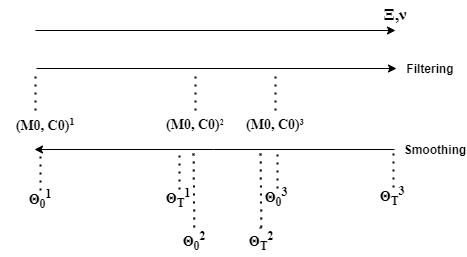}
    \caption{Multiple Time Series}
    \label{fig:concurrent time series}
\end{figure}

The following algorithms defines our approach more formally. 

\begin{algorithm}[H]
    \DontPrintSemicolon
    \KwIn{$\eta, F, G, \gamma, W, M_0, C_0, \Xi_0, \nu_0, N, K$}
    % \KwOut{$A, R, f, q, S, e, M, C, \Xi, \nu$}
    $t = 0$ \\
    $ A, R, f, q, S, e, M, C, \Xi, \nu \gets \emptyset$ \\
    \For{$k = 0 \to K$}{
        $M_{t+k} \gets M_{0}(k)$ \\
        $C_{t+k} \gets C_{0}(k)$ \\
        \For{$i = 0 \to N(k)$}{
            \tcp{Prior}
            $A_t \gets G_t M_{t+k}$ \\
            $R_t \gets G_t C_{t+k} G_t^{T} + W_t$ \\
            \BlankLine
            \tcp{One-step ahead forecast}
            $f_t \gets A_t^{T} F_t$ \\
            $q_t \gets F_t^{T} R_t F_t + \gamma_t$ \\
            \BlankLine
            \tcp{Posterior}
            $S_t \gets R_t F_t / q_t$ \\
            $e_t \gets \eta_t - f_t$ \\
            $M_{t+k+1} \gets A_t + S_t e^T$ \\
            $C_{t+k+1} \gets R_t - q_t S_t S_t^T$ \\
            $\nu_{t+1} \gets \nu_t + 1$ \\
            $\Xi_{t+1} \gets \Xi_t + (e_t e_t^T) / q_t$ \\
            \BlankLine
            $t \gets t + 1$
        }
    }
    \KwRet{$A, R, f, q, S, e, M, C, \Xi, \nu$}
    \caption{Filtering for Multiple Time Series}
\end{algorithm}

\begin{algorithm}[H]
    \DontPrintSemicolon
    \KwIn{$\Xi, \nu, K, N, C, M, G, R, A, seed$}

    $rng \gets \text{random}(seed)$\\
    $\Xi_T \gets (\Xi_T + \Xi_T^T) / 2$ \\
    $\Sigma \gets \text{inv\_wishart\_rng}(\nu_T, \Xi_T, rng)$\\
    $\Sigma \gets (\Sigma + \Sigma^T) / 2.0$\\
    \BlankLine
    $t \gets -1$ \\
    $reset\_flag \gets 1$ \\
    $Z, \Theta, \Theta_0 \gets \emptyset$ \\
    \For{$k = K-1 \to 0$}{
        $reset\_flag \gets 1$ \\
        \For{$i = 0 \to N(k)$}{
            \If{$reset\_flag == 1$}{
                $C_{T+k-t-1} \gets (C_{T+k-t-1} +C_{T+k-t-1}^T) / 2.0$ \\
                $\theta_{T-t-2} \gets \text{matrix\_normal\_rng}(M_{T+k-t-1}, C_{T+k-t-1}, \Sigma, rng)$ \\
                $reset\_flag \gets 0$
            }
            \Else {
                $G_t \gets G_{T-t-1}$ \\
                $Z \gets C_{T+k-t-1} G_t^T R_{T-t-1}^{-1}$ \\
                $C_{T+k-t-1} \gets C_{T+k-t-1} - Z  R_{T-t-1} Z^T$ \\
                $C_{T+k-t-1} \gets (C_{T+k-t-1} + C_{T+k-t-1}^T) / 2.0$ \\
                $M_{T+k-t-1}\gets M_{T+k-t-1} + Z (\theta_{T-t-1} - A_{T-t-1})$ \\
                $\theta_{T-t-2} \gets \text{matrix\_normal\_rng}(M_{T+k-t-1}, C_{T+k-t-1}, \Sigma, rng)$ \\
                \If{$i == N(k) - 1$}{
                    $G_t \gets G_{T-t-2}$ \\
                    $Z \gets C_{T+k-t-2} G_t^T R_{T-t-2}^{-1}$ \\
                    $C_{T+k-t-2} \gets C_{T+k-t-2} - Z R_{T-t-2} Z^T$ \\
                    $C_{T+k-t-2} \gets (C_{T+k-t-2} + C_{T+k-t-2}^T) / 2.0$ \\
                    $M_{T+k-t-2} \gets M_{T+k-t-2} + Z (\theta_{T-t-2} - A_{T-t-2})$ \\
                    $\theta_0(k) \gets \text{matrix\_normal\_rng}(M_{T+k-t-2}, C_{T+k-t-2}, \Sigma, rng)$
                }
            }
        }
    }
    \KwRet{$\Theta,\Theta_0,\Sigma$}

    \caption{Smoothing for Mutiple Time Series}
\end{algorithm}

\newpage
\section{HYPERPARAMETER INFERENCE}\label{section: hyper-infer}

In Section~\ref{subsection: hyperparameter} of the main text, we extend our MLN-DLM model by placing a prior on the hyperparameter  \(W=\text{diag}(w_{1}, \dots, w_{Q})\) to make it learnable and introduce a simple Gibbs sampling approach for inference. This is achieved by sampling from the posterior \(p(\Theta, \Sigma, \eta, w_{1}, \dots, w_{Q} \mid Y)\) using Gibbs updates. Below, we expand on this and provide a detailed explanation of the process.

In main text, we outline our approach to sampling the conditional posterior  \(p(\Theta, \Sigma, \eta \mid  Y, W)\). We now develop the complementary Gibbs steps for \(p(w_{1}, \dots, w_{Q} \mid  Y, \Theta, \eta, \Sigma)\). Note that, by conditional independence, this reduces to \(p(w_{1}, \dots, w_{Q} \mid \Theta, \Sigma)\).

In terms of the conditional \(p(w_{1}, \dots, w_{Q} \mid  \Theta, \Sigma)\), the relevant parts of the MLN-DLM model are
% TODO Manan: again, equations are part of a sentence, they need punctuation
\begin{align*}
    \Theta_t &= G_t\Theta_{t-1} + \Omega_t, \quad \Omega_t \sim \mathcal{N}\left(0, W, \Sigma \right) \\
    W &= \begin{bmatrix}
        w_1 & \cdots & 0 \\
        \vdots & \ddots & \vdots \\
        0 & \cdots & w_Q
        \end{bmatrix} \\
    w_q &\sim \text{InvGamma}(a_q,b_q).
\end{align*}

Letting \(L=\text{Cholesky}(\Sigma^{-1})\), and letting \(\Omega_{t}^{*}=\Omega_{t}L\) this model can be reparameterized as
\begin{align*}
    \Omega^{*}_{ti} &\sim \mathcal{N}\left(0, W \right)\\ 
    W &= \begin{bmatrix}
            w_1 & \cdots & 0 \\
            \vdots & \ddots & \vdots \\
            0 & \cdots & w_Q
        \end{bmatrix} \\
    w_q &\sim \text{InvGamma}(a_q,b_q).
\end{align*}

A blocked Gibbs step can be used to update each \(w_{1}, \dots, w_{Q}\). Each is updated as a standard, conjugate normal-inverse gamma model with posterior
\(p(w_q \mid  \Omega_{q}^{*})=\text{InvGamma}(a_{q}^{*}, b_{q}^{*})\).
The parameters of the posterior are given by 
\begin{align*}
    a_{q}^{*} &= a_{q} + T(D-1) \\ 
    b_{q}^{*} &= \frac{1}{a_{q}^{*}}\left(a_{q}b_{q}^{2} + \sum_{t,i}\left[\Omega^{*}_{qti}- \overline{\Omega_{q}^{*}}\right]^{2}\right)
\end{align*}
where \(\overline{\Omega_{q}^{*}}\) denotes the sample mean of the \(T \times (D-1)\) elements in the matrix \(\Omega_{q}^{*}=\left[\Omega^{*}_{q1}, \dots, \Omega^{*}_{qT}\right]\). 

\newpage
\section{SIMULATIONS}\label{section: simulated-data}

In this section, we give the full details of the data simulation procedure briefly mentioned in the main text's Section~\ref{subsection: simulation}. Also, we provide the results of uncertainty quantification when number of multinomial categories \(D = 10\) with \(T = 300\) to show that our conclusions on the smaller simulation hold in higher dimensions.

\subsection {Data Simulation Model}
We use the MLN-DLM model with the following specified priors to generate simulated data:
\begin{align*}
        Y_{.t} &\sim \mathrm{Multinomial}(\pi_{.t}) \\
        \pi_{.t} &= \phi^{-1}(\eta_{.t}) \\
        \eta_t^{T} &= \Theta_t + v_t^{T} \quad \quad\quad v_t \sim \mathcal{N}(0, \Sigma) \\
        \Theta_t &= \Theta_{t-1} + \Omega_t, \quad \Omega_t \sim \mathcal{N}(0, 0.45 ,\Sigma) \\
        \Theta_0 &\sim \mathcal{N} (M_0, C_0, \Sigma) \\
        \Sigma &\sim \emph{IW} (I,D+3) \\
        M_0 &\sim \mathrm{Uniform}(0.1,1) \\
        C_0 &\sim \mathrm{Uniform}(1,1.5).
\end{align*}

Each simulation consists of multiple time series, with 5\% of the time points randomly missing in each series. While each time series has its own state parameters \(\Theta_{t}\), information about other parameters (e.g., \(\Sigma\)) is shared across the series. Every time series contains 100 time points, and the total number of time series depends on the overall number of time points. For example, 1,000 total time points would correspond to 10 time series, each with 100 time points.

\subsection{Considerations during Uncertainty Quantification}
While generating Debiased Multinomial Dirichlet Samples for Uncertainty Quantification in Fenrir, we choose the pseudocount parameter \(\alpha=0.5\) and generate 2000 samples. For Stan, we run 4 chains, each with 4500 iterations, taking the first 1500 as warmup.

% TODO Manan, before submitting, fix so that figures fit nicely under the heading. As it currently stands it looks messy. You can shrink/enlarge figures a bit to make this work or use \newpage where needed0 -- but be wary and don't introduce too much whitespace. Better to subtlety adjust figure sizes to make everything look nice. 
\subsection{Results showcasing the Bias in posterior Mean of \texorpdfstring{\textbf{$\eta$}}{eta}}
\begin{figure}[!h]
    \centering
    \includegraphics[width=0.5\linewidth]{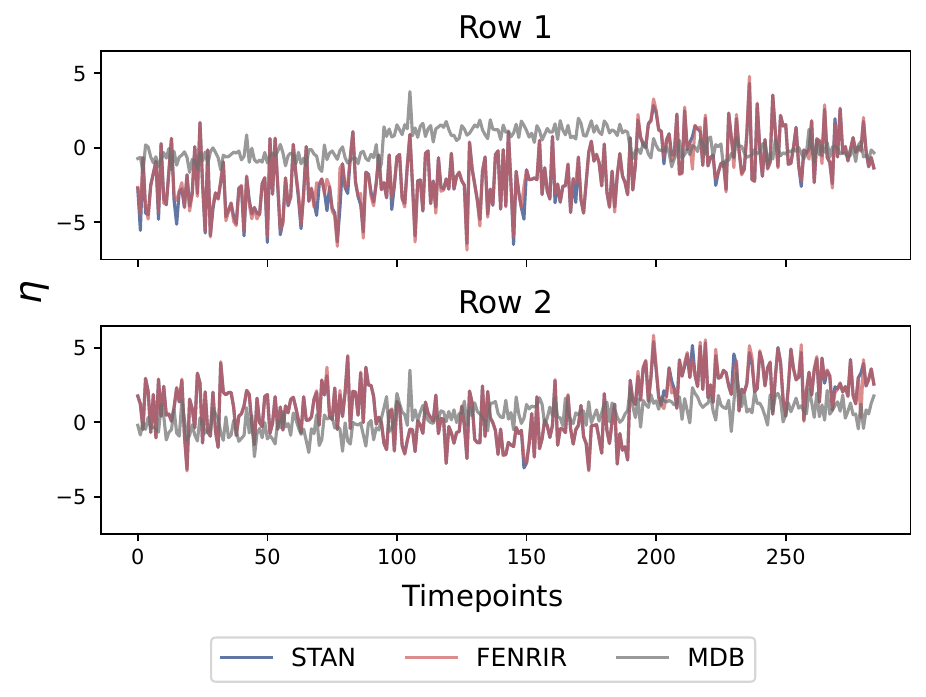}
    \caption{\textbf{Posterior mean for \(\eta\) of MLN-DLM applied to simulated data.} We compared Fenrir's Debiased Multinomial Dirichlet Bootstrap to our Stan implementation and to a version of Fenrir which used the Multinomial Dirichlet Bootstrap on simulated data with \(D=3\) and \(T=300\). Posterior means for each method are plotted in ALR space, resulting in \(D-1\) subplots.}
    \label{fig:Simulated D 10}
\end{figure}

\newpage
\subsection{Results of Uncertainty Quantification for Simulated Data when D = 10 with 1\% Sparsity}
\begin{figure}[!h]
    \centering
    \includegraphics[width=0.8\linewidth]{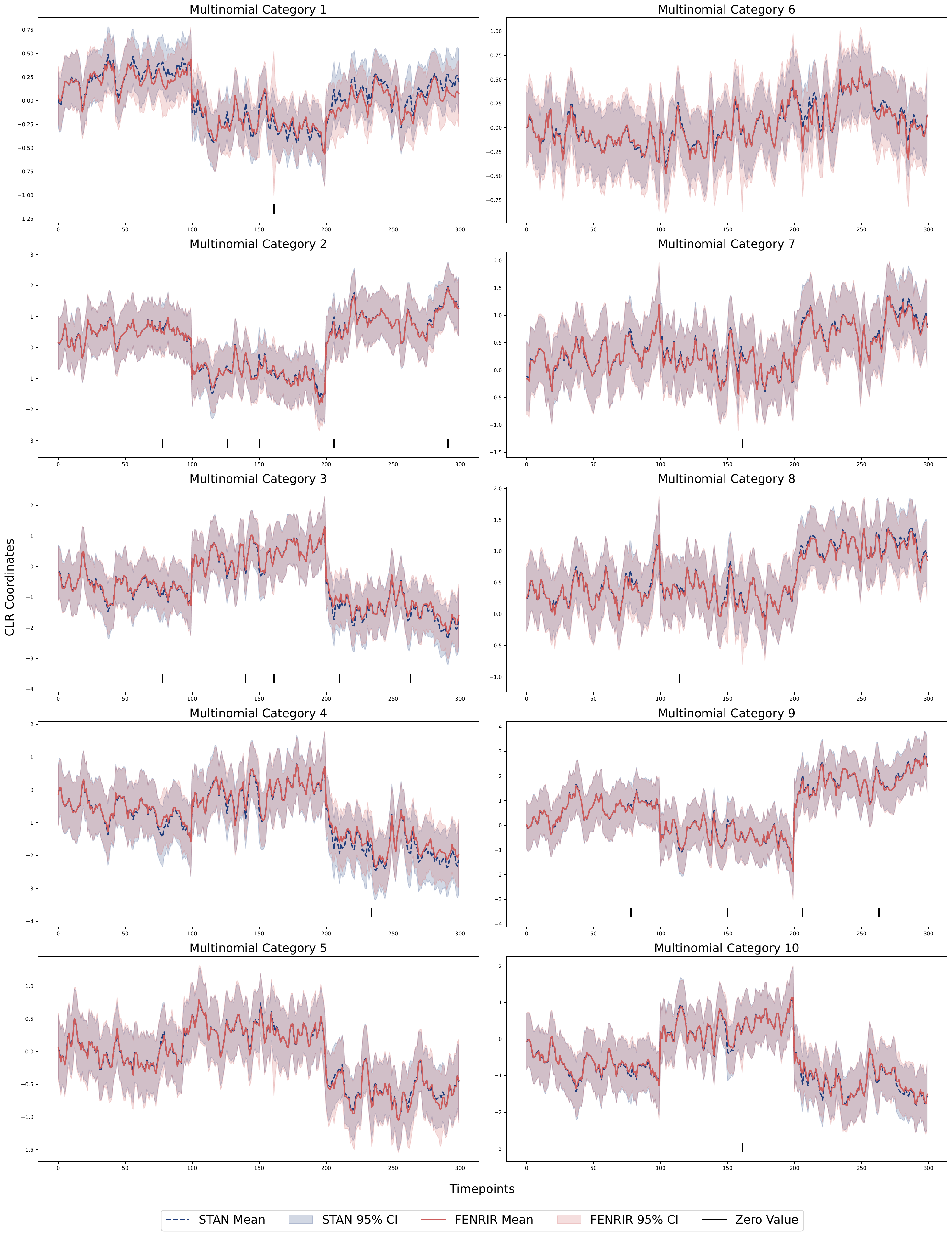}
    \caption{\textbf{Posterior mean and credible intervals for \(\Theta\) of MLN-DLM applied to simulated data.} We compared Fenrir to our Stan implementation on simulated data with \(D=10\) and \(T=300\) with 1\% sparsity (percentage of zero counts in \(Y\)). Posterior means and 95\% credible intervals are plotted in centered log-ratio (CLR) coordinates.}
    \label{fig:Simulated D 10}
\end{figure}

\newpage
\subsection{Results of Uncertainty Quantification for Simulated Data when D = 10 with 9\% Sparsity}
\begin{figure}[!h]
    \centering
    \includegraphics[width=0.8\linewidth]{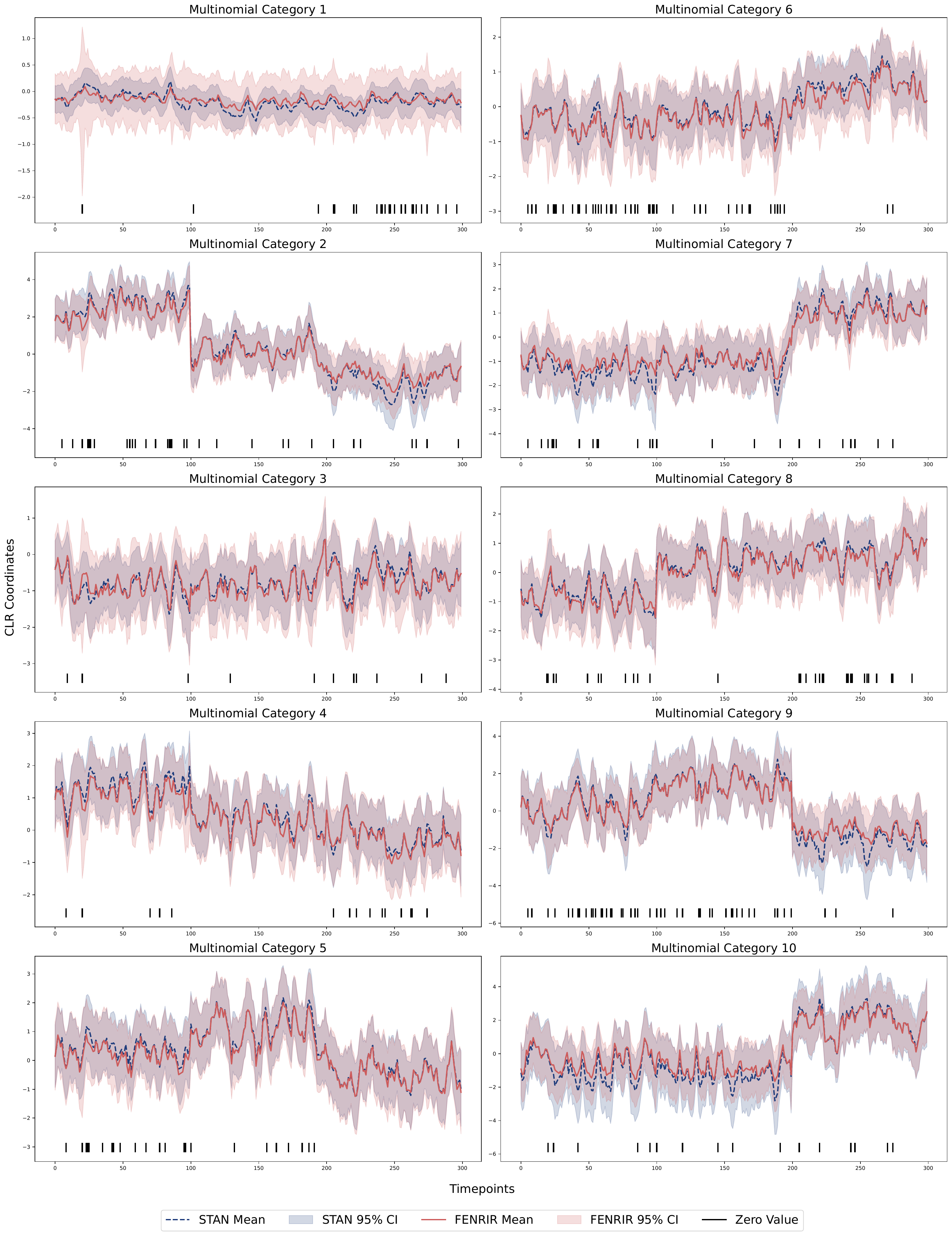}
    \caption{\textbf{Posterior mean and credible intervals for \(\Theta\) of MLN-DLM applied to simulated data.} We compared Fenrir to our Stan implementation on simulated data with \(D=10\) and \(T=300\) with 9\% sparsity (percentage of zero counts in \(Y\)). Posterior means and 95\% credible intervals are plotted in centered log-ratio (CLR) coordinates.}
    \label{fig:Simulated D 10}
\end{figure}

\newpage
\subsection{Results of Uncertainty Quantification for Simulated Data when D = 10 with 20\% Sparsity}
\begin{figure}[!h]
    \centering
    \includegraphics[width=0.8\linewidth]{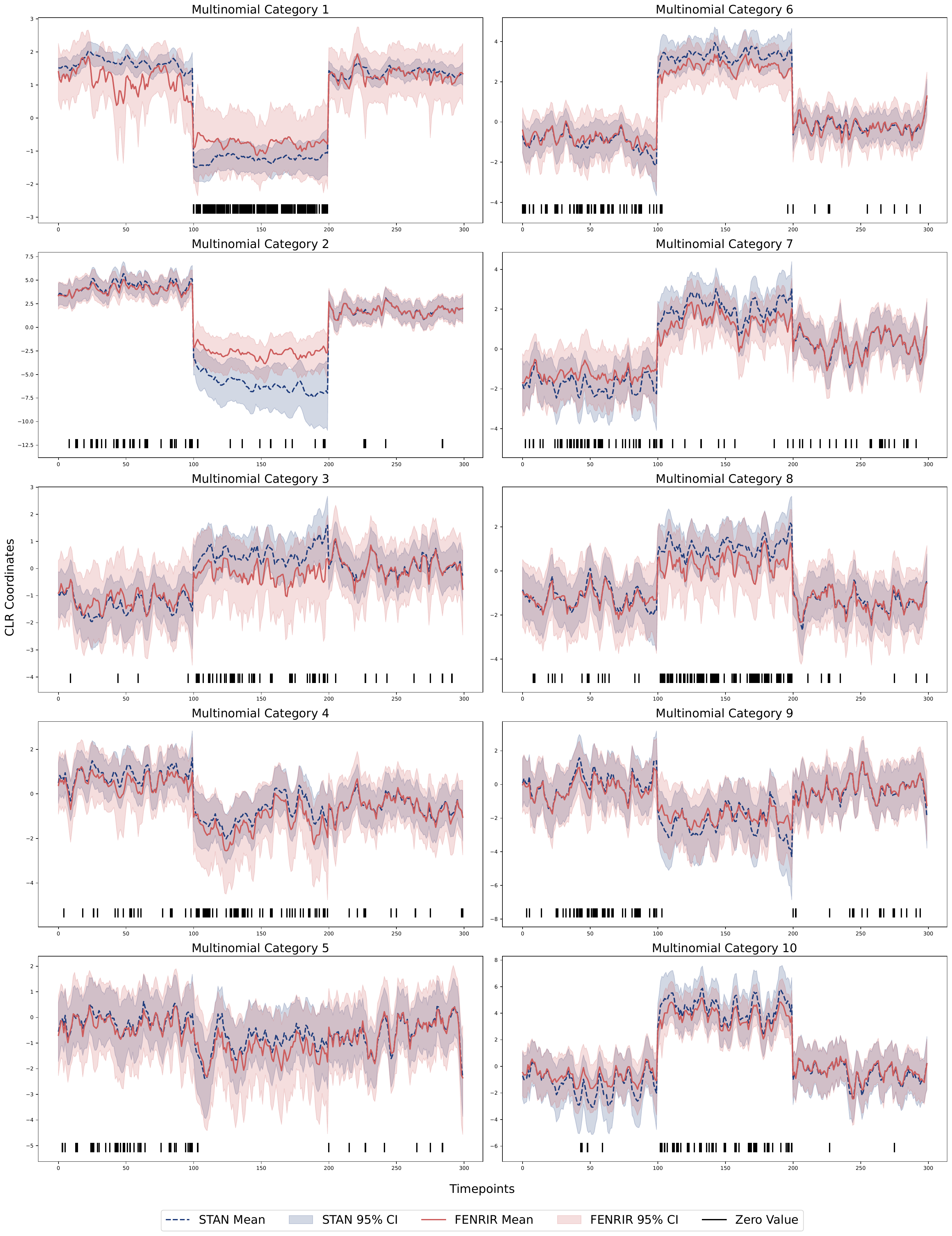}
    \caption{\textbf{Posterior mean and credible intervals for \(\Theta\) of MLN-DLM applied to simulated data.} We compared Fenrir to our Stan implementation on simulated data with \(D=10\) and \(T=300\) with 20\% sparsity (percentage of zero counts in \(Y\)). Posterior means and 95\% credible intervals are plotted in centered log-ratio (CLR) coordinates.}
    \label{fig:Simulated D 10}
\end{figure}

\newpage
\section{ARTIFICIAL GUT MICROBIOME DATA}\label{section: microbiome-data}
We obtained data from the artificial gut study, presented in the main text's Section~\ref{subsection: real-data}, from the R package Fido \href{https://github.com/jsilve24/fido}{(github.com/jsilve24/fido)}. The data is available in that package as the data object \textit{mallard\_family}.  This data contains observations recorded at irregular intervals; the study consisted of both daily and hourly sampling. Our analyses occur at hourly time-scales. As a result, we padded the original data with missing values so the entire series can be represented as hourly.

In the following subsections, we provide details of the MLN-DLM model with prior specifications used to produce the results discussed in Section~\ref{subsection: real-data} of the main text. We also present the posterior estimates of the hyperparameter \(W=\text{diag}(w_{1})\), as well as the posterior inference results for state \(\Theta\) across all taxa and vessels.

\subsection{MLN-DLM Model Specification}
Both Fenrir and Stan fit the model specified below to produce the results here and in the main text.
\begin{align*}
        Y_{.t} &\sim \mathrm{Multinomial}(\pi_{.t})\\
        \pi_{.t} &= \phi^{-1}(\eta_{.t}) \\
        \eta_t^{T} &= \Theta_t + v_t^{T} \quad \quad\quad v \sim \mathcal{N}(0, \Sigma) \\
        \Theta_t &= \Theta_{t-1} + \Omega_t, \quad \Omega_t \sim \mathcal{N}(0, w_1,\Sigma) \\
        \Theta_0 &\sim \mathcal{N} (0, 1, \Sigma) \\
        \Sigma &\sim \emph{IW} (10I,D+3) \\
        w_1 &\sim \mathrm{Inverse Gamma}(30,15).
\end{align*}

\subsection{Maximum A Posteriori (MAP) estimation Results}
\begin{figure}[h]
    \centering
    \includegraphics[width=0.5\linewidth]{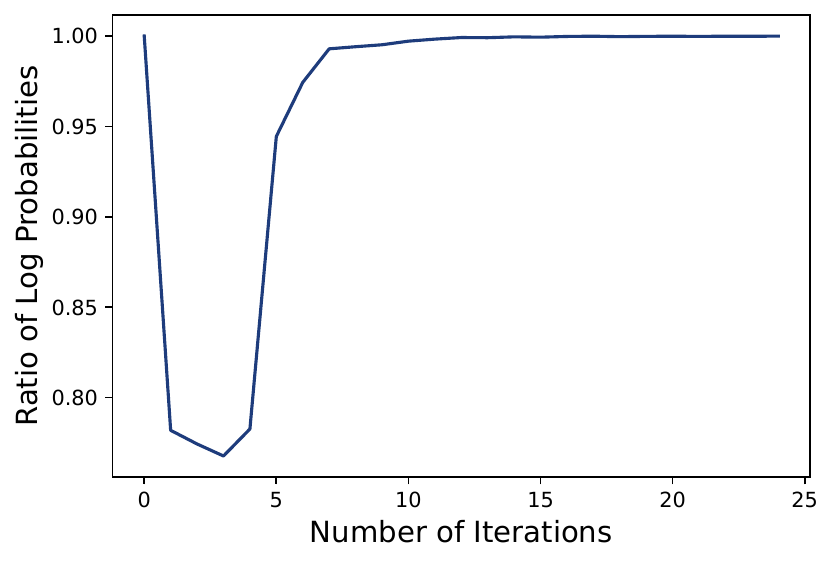}
    \caption{\textbf{Ratio of log probabilities of Fenrir and our Stan implementation for MAP estimation of \(\eta\) applied to Artificial Gut Microbiome Data}}
    \label{fig:mallard logprob}
\end{figure}

\newpage
\subsection{Additional Posterior Inference Results}
\begin{figure}[h]
    \centering
    \begin{minipage}[b]{0.48\linewidth}
        \centering
        \includegraphics[width=\linewidth]{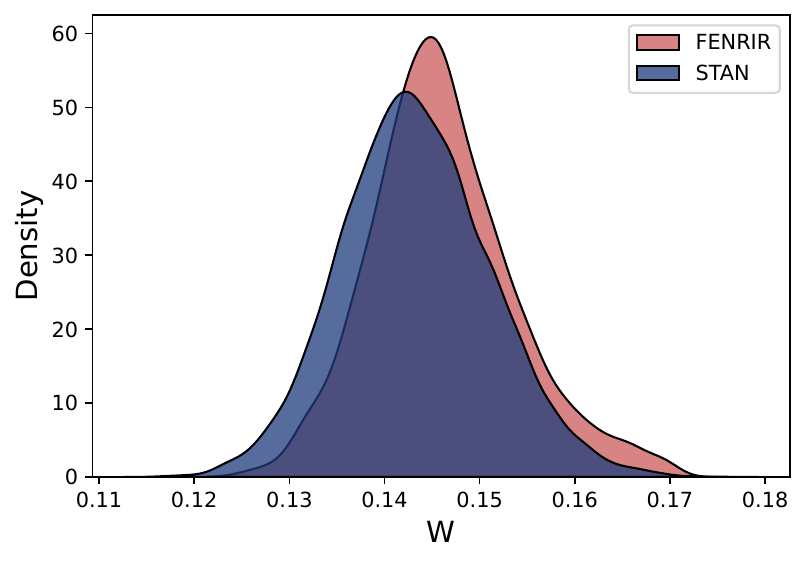}
        \label{fig:density_W}
    \end{minipage}
    \hfill
    \begin{minipage}[b]{0.48\linewidth}
        \centering
        \includegraphics[width=\linewidth]{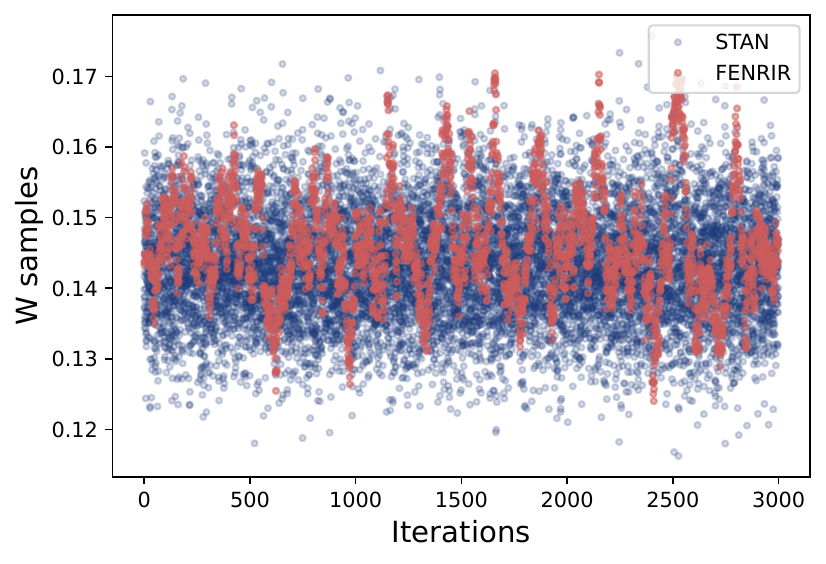}
        \label{fig:traceplot_W}
    \end{minipage}
    \caption{\textbf{Posterior Estimates for hyperparameter \(W\) from Fenrir-based Gibbs sampler and Stan.} The left plot shows the density of posterior samples for \(W\) and the right plot  illustrates the trace of \(W\) samples across 3000 iterations.}
    \label{fig:W_plots}
\end{figure}

\begin{figure}[!h]
    \centering
    \includegraphics[width=0.95\linewidth]{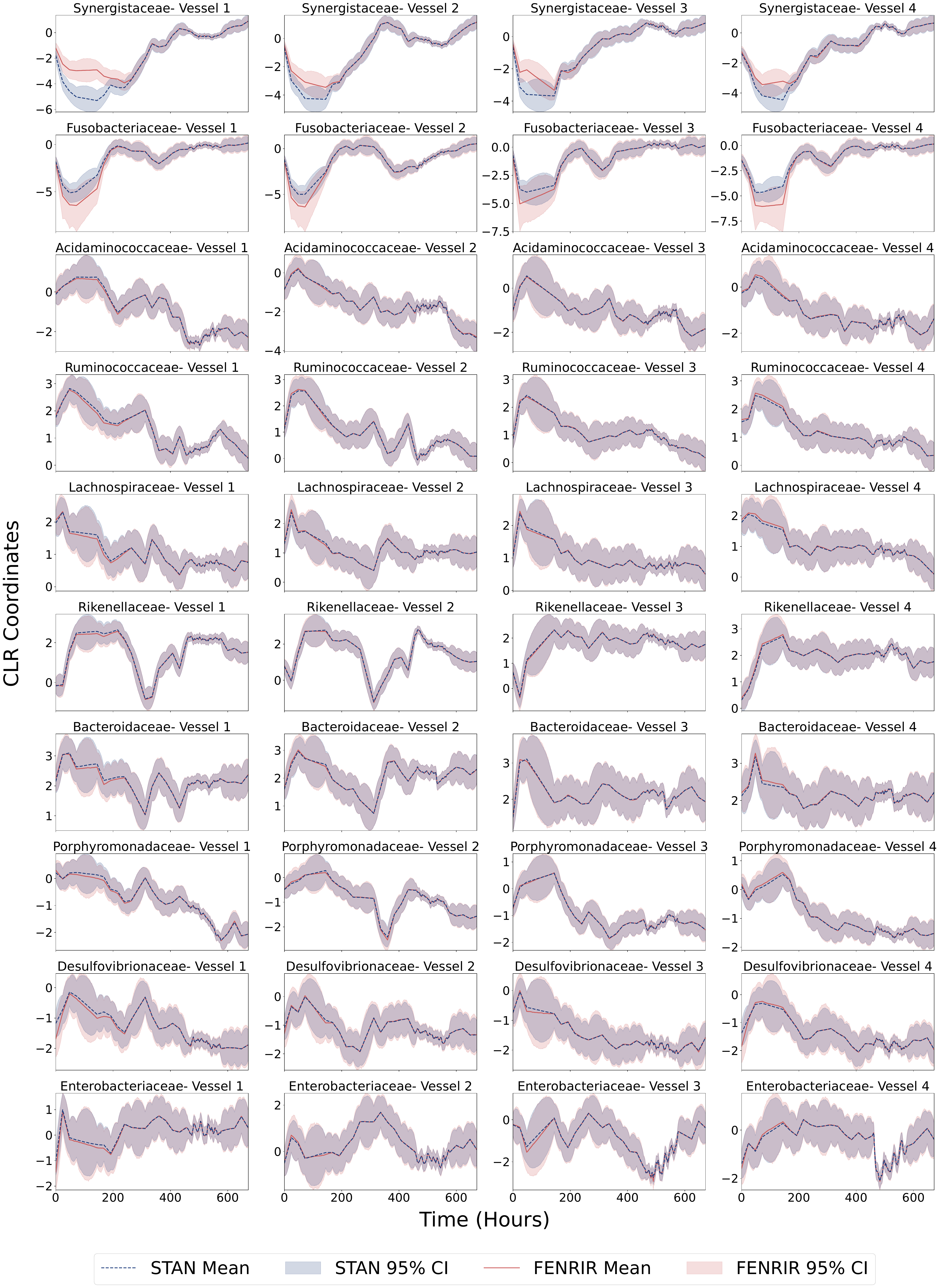}
    \caption{\textbf{Posterior mean and credible intervals for state \(\Theta\) of MLN-DLM applied to Artificial Gut Microbiome Data.} We compared Fenrir to our Stan implementation. Posterior means and 95\% credible intervals are shown in centered log-ratio (CLR) coordiantes for each of the 10 taxa in each of the 4 vessels.}
    \label{fig:Mallard Full}
\end{figure}

\clearpage
\section{LOCAL TREND MLN-DLM MODEL}\label{section: local-trend}

In this section, we aim to demonstrate that Fenrir can model more complex state-space dynamics, such as incorporating a local trend or velocity term. We achieve this by adjusting the MLN-DLM model described in Section~\ref{section: proposed-method} of the main text.

\begin{align*}
    Y_{.t} &\sim \mathrm{Multinomial}(\pi_{.t}) \\
    \pi_{.t} &= \phi^{-1}(\eta_{.t}) \\
    \eta_t^{T} &= F_{t}^{T} \begin{bmatrix} \Theta_t \\ \alpha_t \end{bmatrix} + v_t^{T} \quad \quad\quad v_t \sim \mathcal{N}(0, \gamma_t\Sigma) \\
    \begin{bmatrix} \Theta_t \\ \alpha_t \end{bmatrix} &= G_t\begin{bmatrix} \Theta_{t-1} \\ \alpha_{t-1} \end{bmatrix} + \begin{bmatrix} \Omega_{\Theta t} \\ \Omega_{\alpha t} \end{bmatrix}, \quad \begin{bmatrix} \Omega_{\Theta t} \\ \Omega_{\alpha t} \end{bmatrix} \sim \mathcal{N}\left(0, \begin{bmatrix} w_{\Theta} & 0 \\ 0 & w_{\alpha} \end{bmatrix} ,\Sigma\right) \\
    \begin{bmatrix} \Theta_0 \\ \alpha_0 \end{bmatrix} &\sim \mathcal{N} (M_0, C_0, \Sigma) \\
    \Sigma &\sim \emph{IW} (\Xi,\nu).
\end{align*}
We can extend the above model by including a prior over \(w_{\Theta}\) and \(w_{\alpha}\):
\begin{align*}
    w_{\Theta} &\sim \mathrm{InvGamma}(a,b) \nonumber \\
    w_{\alpha} &\sim \mathrm{InvGamma}(c,d) \nonumber.
\end{align*}

We follow similar steps as in Section~\ref{subsection: hyperparameter} of the main text to derive the posterior conditionals and set up our naive Fenrir-based Gibbs sampler for the model above. Sampling from \(p(\Theta, \Sigma, \eta \mid  Y, W)\) remains the same as discussed in the main text, while \(p(w_\Theta, w_\alpha \mid \Theta, \Sigma)\) is derived below.

Letting \(L=\text{Cholesky}(\Sigma^{-1})\), and letting \(\begin{bmatrix} \Omega_{\Theta t} \\ \Omega_{\alpha t} \end{bmatrix}^{*}=\begin{bmatrix} \Omega_{\Theta t} \\ \Omega_{\alpha t} \end{bmatrix}L\), this model can be reparameterized as
\begin{align*}
    \begin{bmatrix} \Omega_{\Theta ti} \\ \Omega_{\alpha ti} \end{bmatrix}^{*} &\sim \mathcal{N}\left(0, \begin{bmatrix} w_{\Theta} & 0 \\ 0 & w_{\alpha} \end{bmatrix}\right) \\ 
    w_{\Theta} &\sim \mathrm{InvGamma}(a,b) \\
    w_{\alpha} &\sim \mathrm{InvGamma}(c,d).
\end{align*}

Since \(w_{\Theta}\) and \(w_{\alpha}\) are both independent, we can simply calculate the respective posteriors. 
\begin{align*}
    p(w_{\Theta} \mid  \Omega_{\Theta}^{*}) &=\text{InvGamma}(a^{*}, b^{*}) \\
    p(w_{\alpha} \mid  \Omega_{\alpha}^{*}) &=\text{InvGamma}(c^{*}, d^{*}) \\
    a^{*} &= a + T(D-1) \\ 
    b^{*} &= \frac{1}{a^{*}}\left(ab^{2} + \sum_{t,i}\left[\Omega^{*}_{\Theta ti}- \overline{\Omega^{*}_{\Theta}}\right]^{2}\right)\\
    c^{*} &= c + T(D-1) \\ 
    d^{*} &= \frac{1}{c^{*}}\left(cd^{2} + \sum_{t,i}\left[\Omega^{*}_{\alpha ti}- \overline{\Omega^{*}_{\alpha}}\right]^{2}\right)
\end{align*}
where \(\overline{\Omega^{*}}\) denotes the sample mean of the \(T \times (D-1)\) elements in the matrix \(\Omega^{*}=\left[\Omega^{*}_{1}, \dots, \Omega^{*}_{T}\right]\). 

\newpage
We specify the model defined above with the following priors and fit it to the same artificial gut microbiome data discussed in Section~\ref{subsection: real-data} of the main text:
\begin{align*}
    Y_{.t} &\sim \mathrm{Multinomial}(\pi_{.t}) \nonumber\\
    \pi_{.t} &= \phi^{-1}(\eta_{.t}) \nonumber\\
    \eta_t^{T} &= \begin{bmatrix} 1 & 0 \end{bmatrix} \begin{bmatrix} \Theta_t \\ \alpha_t \end{bmatrix} + v_t^{T} \quad \quad\quad v_t \sim \mathcal{N}(0, \Sigma) \\
    \begin{bmatrix} \Theta_t \\ \alpha_t \end{bmatrix} &= \begin{bmatrix} 1 & 1 \\ 0 & 0.9 \end{bmatrix} \begin{bmatrix} \Theta_{t-1} \\ \alpha_{t-1} \end{bmatrix} + \Omega_t, \quad \Omega_t \sim \mathcal{N}\left(0, \begin{bmatrix} w_{\Theta} & 0 \\ 0 & w_{\alpha} \end{bmatrix} ,\Sigma\right) \nonumber\\
    \begin{bmatrix} \Theta_0 \\ \alpha_0 \end{bmatrix} &\sim \mathcal{N} (0, I, \Sigma) \nonumber\\
    \Sigma &\sim \emph{IW} (10I,D + 3) \nonumber \\
    w_{\Theta} &\sim \mathrm{InvGamma}(30,15) \nonumber \\
    w_{\alpha} &\sim \mathrm{InvGamma}(30,8). \nonumber
\end{align*}

The figures below show the posterior inference of state \(\Theta\) and state \(\alpha\) across all taxa and all four vessels of the artificial gut microbiome data producing nearly identical posterior estimates for both methods. Regarding the posterior inference of hyperparameter \(W\), our Gibbs sampler produced 3000 samples in 1.4 hours, while Stan took 12 hours to generate the same number. The posterior estimates of \(w_{\Theta}\) and \(w_{\alpha}\) were identical, with Fenrir estimating a posterior mean and 95\% credible interval of 0.12 (0.104–0.135) for \(w_{\Theta}\) and 0.02 (0.019–0.022) for \(w_{\alpha}\), and Stan estimating a posterior mean and 95\% credible interval of 0.118 (0.104–0.135) for \(w_{\theta}\) and 0.02 (0.019–0.022) for \(w_{\alpha}\). Again even with a simple, naive Gibbs sampler, our approach produced approximately 0.58 effective samples per second for both state \(\Theta\) and \(\alpha\), compared to Stan’s 0.26 and 0.27, respectively, for the two states. We conclude that our approach is approximately twice as fast as Stan with only minimal error in posterior estimates even when modeling complex state-space dynamics.

\newpage
\begin{figure}[!h]
    \centering
    \includegraphics[width=0.95\linewidth]{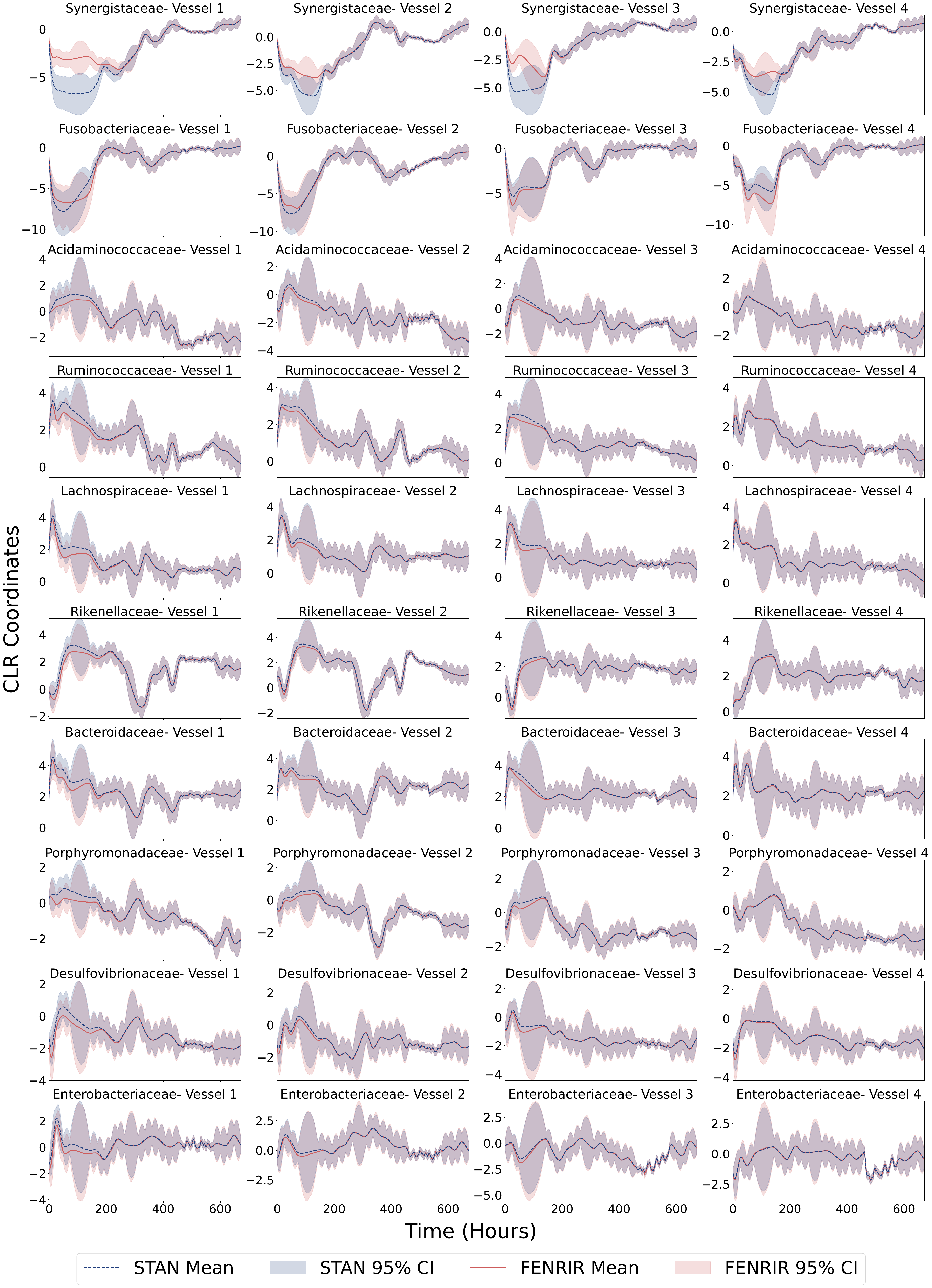}
    \caption{\textbf{Posterior mean and credible intervals for state \(\Theta\) of local trend MLN-DLM model applied to Artificial Gut Microbiome Data.} We compared Fenrir to our Stan implementation. Posterior means and 95\% credible intervals are shown in centered log-ratio (CLR) coordiantes for each of the 10 taxa in each of the 4 vessels.}
    \label{fig:Mallard Vel Theta}
\end{figure}

\newpage
\begin{figure}[!h]
    \centering
    \includegraphics[width=0.95\linewidth]{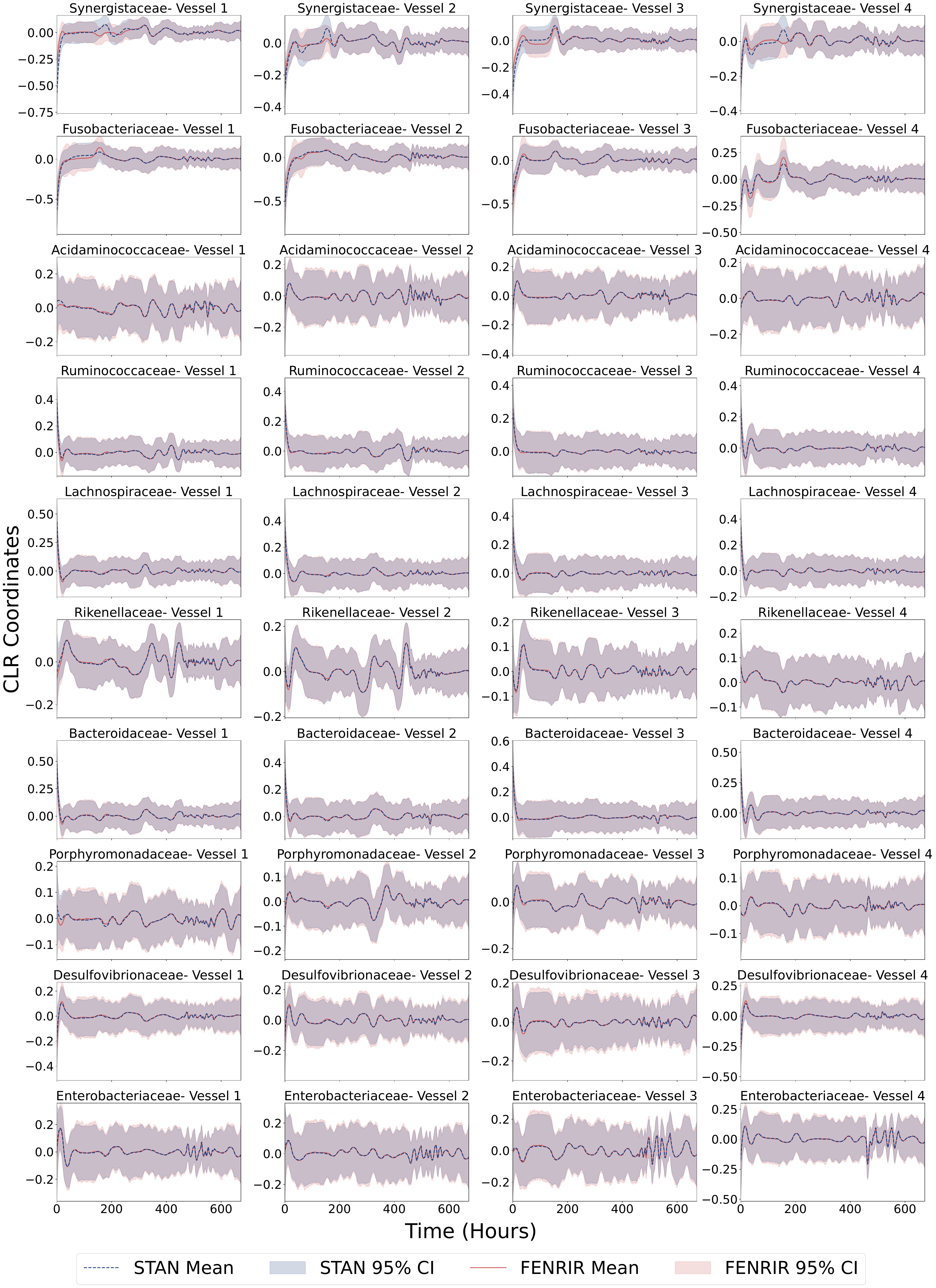}
    \caption{\textbf{Posterior mean and credible intervals for state \(\alpha\) of local trend MLN-DLM model applied to Artificial Gut Microbiome Data.} We compared Fenrir to our Stan implementation. Posterior means and 95\% credible intervals are shown in centered log-ratio (CLR) coordiantes for each of the 10 taxa in each of the 4 vessels.}
    \label{fig:Mallard Vel ALpha}
\end{figure}


\newpage
\clearpage
\begin{thebibliography}{}

\bibitem[{\"A}ij{\"o} et~al., 2018]{aijo2018temporal}
{\"A}ij{\"o}, T., M{\"u}ller, C.~L., and Bonneau, R. (2018).
\newblock Temporal probabilistic modeling of bacterial compositions derived from 16s rrna sequencing.
\newblock {\em Bioinformatics}, 34(3):372--380.

\bibitem[Aitchison, 1982]{aitchison1982statistical}
Aitchison, J. (1982).
\newblock The statistical analysis of compositional data.
\newblock {\em Journal of the Royal Statistical Society: Series B (Methodological)}, 44(2):139--160.

\bibitem[Aitchison and Shen, 1980]{aitchison1980logistic}
Aitchison, J. and Shen, S.~M. (1980).
\newblock Logistic-normal distributions: Some properties and uses.
\newblock {\em Biometrika}, 67(2):261--272.

\bibitem[Cargnoni et~al., 1997]{cargnoni1997bayesian}
Cargnoni, C., M{\"u}ller, P., and West, M. (1997).
\newblock Bayesian forecasting of multinomial time series through conditionally gaussian dynamic models.
\newblock {\em Journal of the American Statistical Association}, 92(438):640--647.

\bibitem[Carpenter et~al., 2017]{carpenter2017stan}
Carpenter, B., Gelman, A., Hoffman, M.~D., Lee, D., Goodrich, B., Betancourt, M., Brubaker, M.~A., Guo, J., Li, P., and Riddell, A. (2017).
\newblock Stan: A probabilistic programming language.
\newblock {\em Journal of statistical software}, 76.

\bibitem[Espinoza et~al., 2020]{espinoza2020applications}
Espinoza, J.~L., Shah, N., Singh, S., Nelson, K.~E., and Dupont, C.~L. (2020).
\newblock Applications of weighted association networks applied to compositional data in biology.
\newblock {\em Environmental Microbiology}, 22(8):3020--3038.

\bibitem[Fernandes et~al., 2014]{fernandes2014unifying}
Fernandes, A.~D., Reid, J.~N., Macklaim, J.~M., McMurrough, T.~A., Edgell, D.~R., and Gloor, G.~B. (2014).
\newblock Unifying the analysis of high-throughput sequencing datasets: characterizing {RNA-seq}, {16S} {rRNA} gene sequencing and selective growth experiments by compositional data analysis.
\newblock {\em Microbiome}, 2(1):1--13.

\bibitem[Fokianos and Kedem, 2003]{fokianos2003categoricaltimeseries}
Fokianos, K. and Kedem, B. (2003).
\newblock Regression theory for categorical time series.
\newblock {\em Statistical Science}, 18(3):357--376.

\bibitem[Friedman and Alm, 2012]{friedman2012sparcc}
Friedman, J. and Alm, E.~J. (2012).
\newblock Inferring correlation networks from genomic survey data.
\newblock {\em PLoS Computational Biology}, 8(9):e1002687.

\bibitem[Glynn et~al., 2019]{glynn2019bayesian}
Glynn, C., Tokdar, S.~T., Howard, B., and Banks, D.~L. (2019).
\newblock Bayesian {Analysis} of {Dynamic} {Linear} {Topic} {Models}.
\newblock {\em Bayesian Analysis}, 14(1).

\bibitem[Grantham et~al., 2020]{grantham2020mimix}
Grantham, N.~S., Guan, Y., Reich, B.~J., Borer, E.~T., and Gross, K. (2020).
\newblock Mimix: A bayesian mixed-effects model for microbiome data from designed experiments.
\newblock {\em Journal of the American Statistical Association}, 115(530):599--609.

\bibitem[Joseph et~al., 2020]{joseph2020efficient}
Joseph, T.~A., Pasarkar, A.~P., and Pe’er, I. (2020).
\newblock Efficient and accurate inference of mixed microbial population trajectories from longitudinal count data.
\newblock {\em Cell Systems}, 10(6):463--469.

\bibitem[Linderman et~al., 2015]{linderman2015dependent}
Linderman, S., Johnson, M.~J., and Adams, R.~P. (2015).
\newblock Dependent multinomial models made easy: Stick-breaking with the p{\'o}lya-gamma augmentation.
\newblock {\em Advances in neural information processing systems}, 28.

\bibitem[Mosimann, 1962]{mosimann1962compound}
Mosimann, J.~E. (1962).
\newblock On the compound multinomial distribution, the multivariate $\beta$-distribution, and correlations among proportions.
\newblock {\em Biometrika}, 49(1/2):65--82.

\bibitem[Murray and Adams, 2010]{murray2010slicegauss}
Murray, I. and Adams, R.~P. (2010).
\newblock Slice sampling covariance hyperparameters of latent {Gaussian} models.
\newblock {\em Advances in neural information processing systems}, 23.

\bibitem[Nixon et~al., 2024]{nixon2024aldex2}
Nixon, M.~P., Gloor, G.~B., and Silverman, J.~D. (2024).
\newblock Beyond {Normalization}: {Incorporating} {Scale} {Uncertainty} in {Microbiome} and {Gene} {Expression} {Analysis}.
\newblock {\em bioRxiv}.

\bibitem[Nixon et~al., 2023]{nixon2023sri}
Nixon, M.~P., Letourneau, J., David, L.~A., Lazar, N.~A., Mukherjee, S., and Silverman, J.~D. (2023).
\newblock {S}cale {R}eliant {I}nference.
\newblock {\em arXiv}.

\bibitem[Pawlowsky-Glahn et~al., 2015]{pawlowsky2015book}
Pawlowsky-Glahn, V., Egozcue, J.~J., and Tolosana-Delgado, R. (2015).
\newblock {\em Modeling and analysis of compositional data}.
\newblock John Wiley \& Sons.

\bibitem[Polson et~al., 2013]{polson2013bayesian}
Polson, N.~G., Scott, J.~G., and Windle, J. (2013).
\newblock Bayesian inference for logistic models using p{\'o}lya--gamma latent variables.
\newblock {\em Journal of the American statistical Association}, 108(504):1339--1349.

\bibitem[Prado et~al., 2021]{prado2021book}
Prado, R., Ferreira, M. A.~R., and West, M. (2021).
\newblock {\em Time Series: {Modeling}, {Computation}, and {Inference}}.
\newblock Chapman and Hall/CRC, New York, 2 edition.

\bibitem[Qiu et~al., 2023]{rcppnumerical}
Qiu, Y., Balan, S., Beall, M., Sauder, M., Okazaki, N., and Hahn, T. (2023).
\newblock {\em RcppNumerical: 'Rcpp' Integration for Numerical Computing Libraries}.
\newblock R package version 0.6-0.

\bibitem[Silverman et~al., 2018]{silverman2018dynamic}
Silverman, J.~D., Durand, H.~K., Bloom, R.~J., Mukherjee, S., and David, L.~A. (2018).
\newblock Dynamic linear models guide design and analysis of microbiota studies within artificial human guts.
\newblock {\em Microbiome}, 6:1--20.

\bibitem[Silverman et~al., 2022]{silverman2022bayesian}
Silverman, J.~D., Roche, K., Holmes, Z.~C., David, L.~A., and Mukherjee, S. (2022).
\newblock Bayesian multinomial logistic normal models through marginally latent matrix-t processes.
\newblock {\em Journal of Machine Learning Research}, 23(7):1--42.

\bibitem[West and Harrison, 2006]{west2006bayesian}
West, M. and Harrison, J. (2006).
\newblock {\em Bayesian forecasting and dynamic models}.
\newblock Springer Science \& Business Media.

\end{thebibliography}
\end{document}